\documentclass[a4paper,fleqn,usenatbib]{mnras}

\usepackage{newtxtext,newtxmath}

\usepackage[T1]{fontenc}

\usepackage{graphicx}	
\usepackage{amsmath}	

\usepackage{amssymb}	
\usepackage{longtable}
\usepackage{xcolor}
\usepackage{adjustbox}
\usepackage{hyperref}
\usepackage{siunitx}
\usepackage{xspace}
\usepackage{listings}
\usepackage{float}
\DeclareSIUnit\year{yr}
\DeclareSIUnit\mag{mag}

\definecolor{orange}{rgb}{1,0.5,0}
\definecolor{darkgreen}{rgb}{0,0.5,0}
\definecolor{codegreen}{rgb}{0,0.6,0}
\definecolor{codegray}{rgb}{0.5,0.5,0.5}
\definecolor{codepurple}{rgb}{0.58,0,0.82}
\definecolor{backcolour}{rgb}{0.95,0.95,0.92}
\graphicspath{{./}{figures/}}
\newcommand{\winrat}{DWR}
\newcommand{\PIPS}{\texttt{PIPS}\xspace}

\newcommand{\rev}[1]{#1}
\newcommand{\revv}[1]{#1}

\newcommand{\given}{\ | \ }
\lstdefinestyle{mystyle}{
    backgroundcolor=\color{backcolour},   
    commentstyle=\color{codegreen},
    keywordstyle=\color{magenta},
    numberstyle=\tiny\color{codegray},
    stringstyle=\color{codepurple},
    basicstyle=\ttfamily\footnotesize,
    breakatwhitespace=false,         
    breaklines=true,                 
    captionpos=b,                    
    keepspaces=true,                 
    numbersep=5pt,                  
    showspaces=false,                
    showstringspaces=false,
    showtabs=false,                  
    tabsize=2
}
\title[\PIPS I: Fourier-likelihood periodogram \& RR Lyraes]{\texttt{PIPS}, an advanced platform for period detection in time series -- I. Fourier-likelihood periodogram and application to RR Lyrae stars}

\author[Murakami et al.]{Yukei S. Murakami,$^{1,2,3}$\thanks{E-mail: sterling.astro@berkeley.edu}
Connor Jennings,$^{1}$
Andrew M. Hoffman,$^{1,2,4}$
Arjun B. Savel,$^{5}$
\newauthor
James Sunseri,$^{1,2}$
Raphael Baer-Way,$^{1}$
Benjamin E. Stahl,$^{1}$
Ivan Altunin,$^{1}$
Nachiket Girish,$^{2}$
\newauthor
and Alexei V. Filippenko$^{1}$
\\
$^{1}$Department of Astronomy, University of California, Berkeley, CA 94720-3411, USA\\
$^{2}$Department of Physics, University of California, Berkeley, CA 94720-7300, USA\\
$^{3}$Department of Physics and Astronomy, Johns Hopkins University, Baltimore, MD 21218, USA\\
$^{4}$Institute for Astronomy, University of Hawai’i, 2680 Woodlawn Drive, Honolulu, HI 96822, USA\\
$^{5}$Department of Astronomy, University of Maryland, College Park, MD 20742, USA\\
}

\date{Accepted XXX. Received YYY; in original form ZZZ}

\pubyear{2021}

\begin{document}
\label{firstpage}
\pagerange{\pageref{firstpage}--\pageref{lastpage}}
\maketitle

\begin{abstract}
    We describe the \texttt{Period detection and Identification Pipeline Suite} (\PIPS) --- a new, \rev{fast, and statistically robust} platform for period detection and analysis of astrophysical time-series data. 
    \PIPS is an open-source Python package that provides various pre-implemented methods and a customisable framework for automated, robust period measurements with principled uncertainties and statistical significance calculations. In addition to detailing the general algorithm that underlies \PIPS, this paper discusses one of \PIPS' central and novel features, the Fourier-likelihood periodogram, and compares its performance to existing methods. The resulting improved performance implies that one can construct deeper, larger, and more reliable sets of derived properties from various observations, including all-sky surveys. We present a comprehensive validation of \PIPS against artificially generated data, which demonstrates the reliable performance of our algorithm for a class of periodic variable stars (RR Lyrae stars). 
\end{abstract}




    \begin{figure*}
        \centering
        \includegraphics[width=\linewidth]{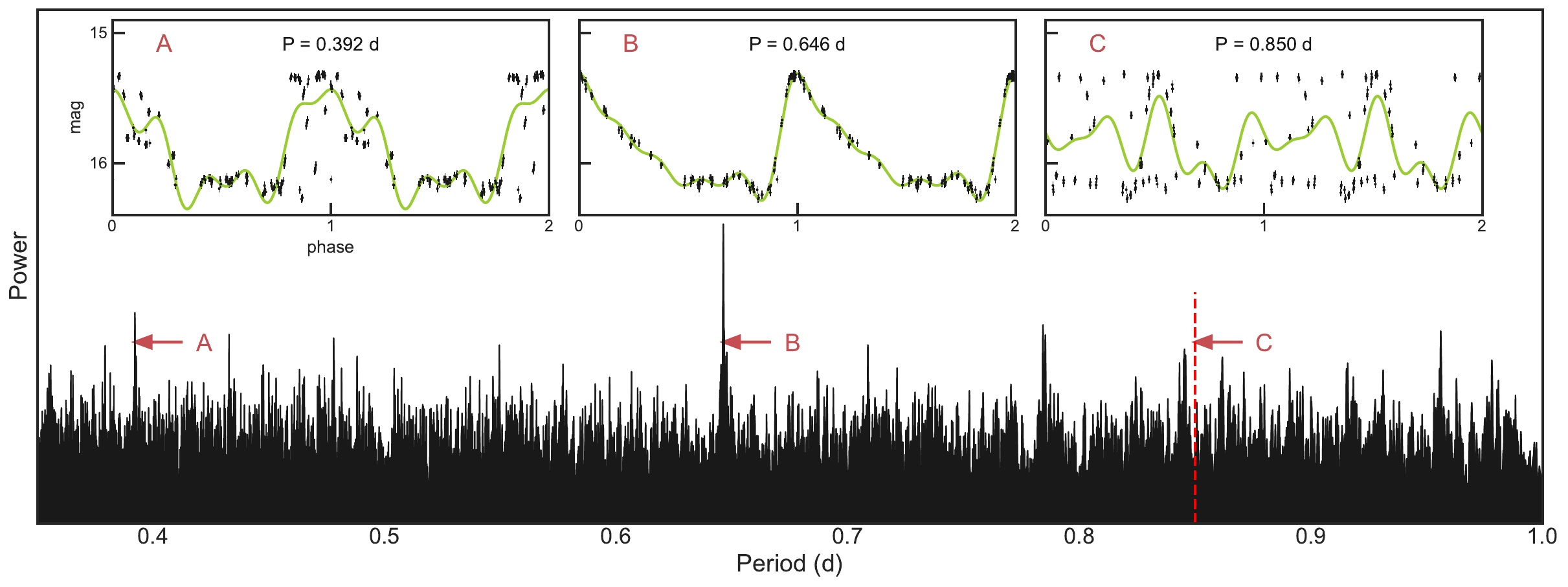}
        \caption{\rev{A visual demonstration of the $\chi^2$ periodogram in which the template function is fitted to phase-folded data \citep[i.e., $V$-band observations of the RRab star dubbed V008 by][]{H&M2020} at each period and the resulting goodness-of-fit measured by $\chi^2$ is translated into the strength spectrum (black). Inset A shows the secondary peak (false signal) in the periodogram and the phase-folded light curve at that period. The data show some structure, but the dispersion is still large around the best-fit curve. Insert B demonstrates the primary peak (true signal), which is in agreement with the value in the literature derived by various independent methods \citep[see][for a detailed comparison]{H&M2020}. Inset C shows a randomly selected location (noise) in the periodogram. The phase-folded data at this period lack any structure and are broadly dispersed.}}
        \label{fig:periodogram_demo}
    \end{figure*}

\section{Introduction} 
    \label{sec:intro}
    Photometric time-series data are highly effective in allowing astronomers to examine the behaviour of distant objects. By studying the pattern, duration, and degree of change in the brightness of such objects as they vary over time, deep insights can be made regarding the astrophysical engines that drive these variations. 
    For instance, gravitational microlensing can be used to study black holes or other significant gravitational sources \citep{Paczynski1986_microlensing}, and the explosions of stars (i.e., supernovae) afford insights into the intense conditions of stellar interiors \citep[e.g.,][and references therein]{Branch_2017_SNbook}. 
    A particularly interesting class of such astrophysical objects is periodic variable stars, which exhibit large, periodic changes in their luminosity. Most commonly discovered are pulsating variable stars \citep[see, e.g.,][for a review]{Percy_2007_VS_Review}.
    For such objects, a host of properties, including metallicity, mass, luminosity, and size are encoded in their brightness variations \citep[e.g.,][]{Leavitt_1912_Ceph_PL,Sandage_Tammnann_1968_CephPCL,Christy_1968_Ceph_th,Jurcsik_Kovacs_1996_RRL_Z}.
    It is, therefore, valuable to determine from a light curve both the pulsation period and ``shape" of brightness variations as a function of time.
    
    The \rev{most straightforward} method for performing such analyses is to conduct a high-cadence observing campaign that covers the entire pulsation period. The period can then be calculated by measuring the consecutive times (i.e., epochs) of light-curve extrema. This method can be found in the literature dating back several centuries, including \cite{Goodricke_1786_DeltaCephei}. However, such high-cadence observations require a great deal of telescope time and are rarely feasible for variable stars.
    
    In many instances, ranging from dedicated observations to modern all-sky surveys \citep[e.g.,][respectively]{Filippenko1981,GaiaDR2_variability}, the data are often sparse and unevenly sampled, obscuring the features of the light curve in the raw time-series data \citep{Lomb1976}. These data also often lack consistent observations of maxima, making it difficult to determine the corresponding period using the simple (high-cadence) method. 
    This data sparsity necessitates a more involved approach for deriving the period that utilises robust mathematical tools to recover the continuous shape and features of the light curve by phase-folding the data \citep{photometry_2007}.

    Many methods have been proposed to perform this task. For observations at a fixed cadence, the discrete Fourier transform (DFT) provides the spectral power in frequency space from which one can estimate the period. Most observations, however, are not obtained with uniform temporal sampling, and thus more advanced tools are required\rev{; for example, the method proposed by \cite{kurtz_dft}}. A practical implementation of least-squares spectral analysis was proposed by \cite{Lomb1976} and \cite{Scargle1982}, and it has since been widely accepted as a standard method (the Lomb-Scargle periodogram; LS periodogram, hereafter) for finding periodic signals. For more specific cases where the target light curve is well known, the method of template fitting (e.g., template Fourier fitting (TFF) described by \citealt{Kovacs_Kupi_2007_TFF}), which finds the period that minimises the deviation of resulting phase-folded light curves from the known shape, can also be effective.
    \rev{Alternative, template-independent approaches have also been proposed, such as phase dispersion minimisation \citep[PDM;][]{Stellingwerf1978_PDM}. While this method has an obvious advantage over the methods introduced above in that it can avoid the possible bias due to the choice of template, it has been shown that PDM is not as sensitive to the signal as the least-squares method owing to the embedded parameterisation in phase binning \citep[][]{Schwarzenberg-Czerny1999_OptimumPeriodSearch}.} \rev{A variety of signal-finding tools implement the methods above, with unique improvements to enhance the reliability and reduce computational costs (e.g., \texttt{SigSpec}, \citealt{sigspec}; \texttt{PERIOD04}, \citealt{coast}).}

    \rev{Unfortunately, least-squares spectral analysis and template fitting are only capable of inferring the period of a light curve, and optimisation or uncertainty estimation of the period and best-fit analytic light curve require separate, additional analysis techniques, such as Markov Chain Monte Carlo (MCMC) or linear/nonlinear regression \citep[e.g.,][]{Eyer2017_GaiaDR1_var}. While these methods can provide reliable and accurate results, they are also extremely sensitive to initial conditions and the range of parameters explored.}

    \rev{Moreover, the method employed to determine the statistical significance of detected signals is not well-established. Thus, improving this aspect of the analysis --- i.e., automatically and statistically informing the selection of detected signals --- is especially important in the current era of large all-sky surveys.}
    
    \rev{In this work, we present an all-in-one, fast, and statistically robust period-determination approach. Our method utilises statistical information that has been embedded in the periodogram but not previously used to simultaneously (i) determine the period, (ii) estimate the uncertainty of the period detection, and (iii) compute the statistical significance of the detected signal without additional analyses. This method, along with other features, is implemented in the publicly available Python package \PIPS} (\texttt{Period \rev{Detection} and Identification Pipeline Suite})\footnote{\label{note:PIPS}PIPS is available at \url{https://github.com/SterlingYM/PIPS}. The basic usage is described in Appendix~\ref{sec:PIPS_usage}.}. We specifically focus herein on its most basic and flexible method to perform period detection with a Fourier-likelihood periodogram, and we also showcase its application to the study of RR Lyrae stars.
    With pre-optimised parameters that are further customisable, \PIPS determines the period and analyses the shape of a light curve simultaneously, minimising template bias and improving overall accuracy and reliability.
    
    \rev{
    We begin our discussion in Sec.~\ref{sec:chi-square-periodogram} by reviewing the concepts of the $\chi^2$ periodogram and several associated issues that necessitate additional analyses. We then propose a solution to said issues in Sec.~\ref{sec:likelihood}, and follow up by describing the design and algorithm of our \texttt{Python} implementation, \PIPS (Sec.~\ref{sec: algorithm}).
    Sec.~\ref{sec: Validation} serves to evaluate the performance of \PIPS using simulated data.
    In addition, we discuss, in Sec.~\ref{sec:additional-modules}, a variety of additional features in \PIPS, such as classification and long-term modulation analysis. Finally, we visit a variety of individual topics related to our work, as well as possible areas of further applications and developments, in Sec.~\ref{sec:discussion}.
    }

    \begin{figure}
        \centering
        \includegraphics[width=\linewidth]{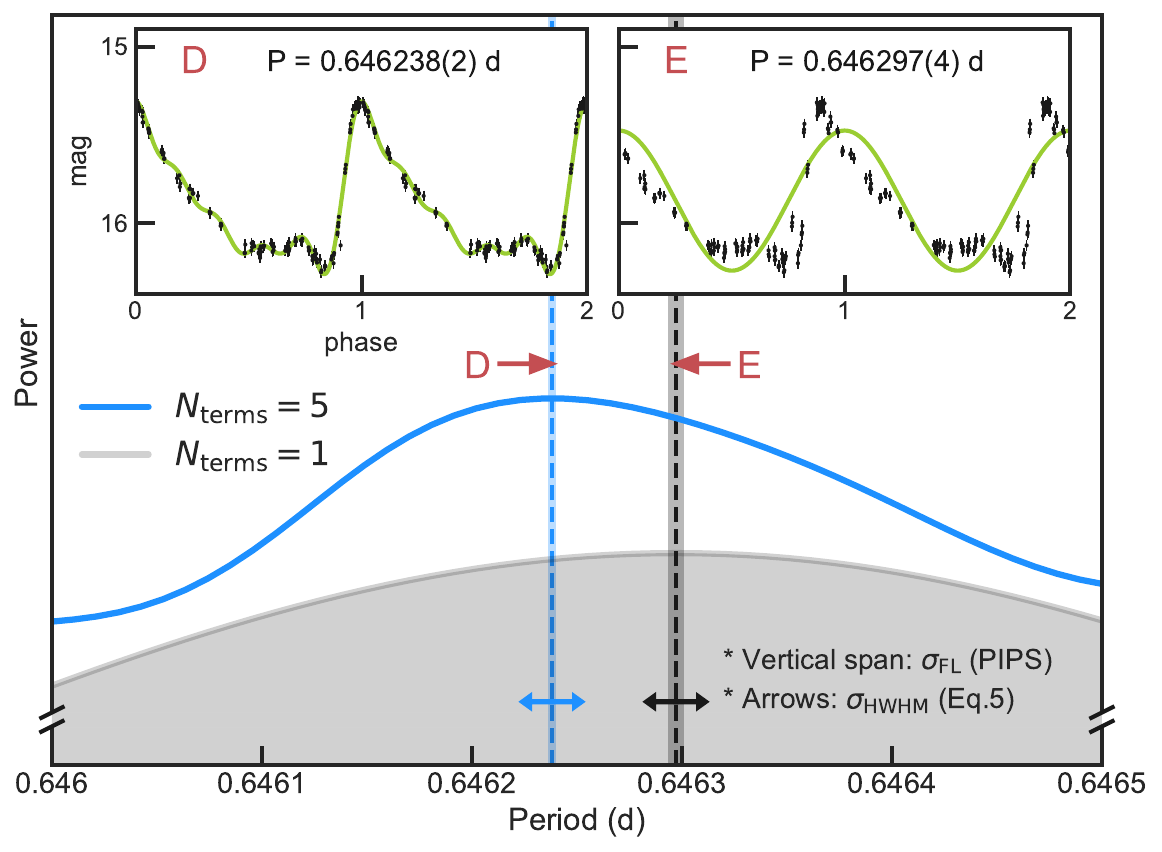}
        \caption{\rev{A visual demonstration of potential issues with the $\chi^2$ periodogram. Two periodograms (blue and grey) generated with different templates (single- and five-term Fourier series) are shown. The difference of peak locations and heights (D and E) illustrate the template bias. The difference in sizes between the HWHM--based uncertainty (Eq.~\ref{eq: stat_uncertainty}) and \PIPS uncertainty (which is approximately equivalent to MCMC-based and linear-regression-based uncertainty values) illustrates the difficulty of directly estimating uncertainty from the periodogram itself.}}
        \label{fig: periodogram_issues}
    \end{figure}

\section{Chi-square periodogram}
    \label{sec:chi-square-periodogram}

    \subsection{Concept}
    \label{sec:periodogram_concept}
        \rev{As we introduce our improvements to period detection with our periodogram process, a solid understanding of the traditional and most commonly used $\chi^2$ interpretation of the periodogram is essential. In this $\chi^2$ framework,} the signal strength in period space is represented by $\chi^2$ values normalised to the best-fit curve fitted at each period. \rev{A visualisation of this concept is provided in Fig.~\ref{fig:periodogram_demo}.}
        As introduced in Sec.~\ref{sec:intro}, the most standard form of $\chi^2$ periodogram is the LS \citep{Lomb1976,Scargle1982}, for which a sinusoidal signal is fitted to the data. This method is a direct application of the discrete Fourier Transform (DFT) for unevenly sampled data, and it has numerous statistical benefits \citep[e.g., calculating false-alarm probability (FAP); for a review, see][]{VanderPlas_2018}.
        
        \rev{More generally, the $\chi^2$ periodogram is created by fitting a selected model function $\mathcal{F}$ to the phase-folded data,
        \begin{equation}
            y_\text{fit}(x; P_\text{test}) = \mathcal{F}(\phi; P_\text{test}, \theta_1,\theta_2, \cdots , \theta_n) \ .
        \end{equation}
        Here, $\phi \equiv (x \mod P_\text{test})$ is the time series data phase-folded at the test period, $P_\text{test}$.
        The variables $\theta_1,\theta_2,\cdots,\theta_n$, are free parameters determined via fitting.}
            
        \rev{This best-fit function serves as a ``template'' for the measurement of the spectral energy density. The goodness of fit can be evaluated with the reduced $\chi^2$ value
        \begin{equation}
            \label{eq: chi-square}
            \sum \chi^2_\nu(P_\mathrm{test}) = \frac{1}{\nu}\sum_j 
            \left(\frac{y_j-  y_\mathrm{fit}(x_j,P_\mathrm{test})}{\sigma_{y_j}}\right)^2\ ,
        \end{equation}
        where $\nu=N-n_\theta$ is the number of degrees of freedom and the subscript $j$ denotes the data point.
        In studies such as those of \cite{bretthorst_1988_stats} or \cite{Palmer2009_chisq_periodogram}, the list of such $\chi^2$ values across the search range is used to construct the periodogram. \cite{VanderPlas_2018} defines the general case of a normalised periodogram as
        \begin{equation}
            \label{eq: periodogram}
            \mathcal{P} = 1 - \frac{\chi^2}{\chi^2_0}\ ,
        \end{equation}
        where the $\chi^2$ values are normalised with the nonvarying $\chi^2$ value $\chi^2_0 = \sum_i (y_i-\overline{y})^2/{\sigma_{y,i}}^2$, which represents the maximum possible $\chi^2$ value for the data. In this formulation of the periodogram, the normalisation $1/\nu$ is canceled.}

    \subsection{Challenges and Current Solutions}
        \rev{
        In applications to modern astronomy, the $\chi^2$ periodogram poses three challenges to users ---
        namely template bias, uncertainty estimation, and significance estimation. Each of these has been discussed in the literature, and several solutions have been proposed.
        }
        
        \subsubsection{Template bias}
        \rev{Despite its popularity, results from an LS periodogram are known to deviate from true periods \citep[e.g., ][]{Schwarzenberg-Czerny1989_AoV}, mainly owing to the fact that LS is a correlation function to a sinusoid. Its detected period can therefore deviate from the true value when the signal is nonsinusoidal. Similarly, other template-fitting methods (e.g., TFF) only guarantee the best match of the data to a template, resulting in performance being dependent upon on the completeness of the template library.}
        \rev{In general, this bias occurs when the light-curve model used to construct the periodogram ($\mathcal{F}$) differs from the model one wishes to use for the final analysis.
        For pulsating variable stars, a multiterm Fourier series is most commonly used for the final (astrophysical) analysis, and using the same model for the periodogram can solve this issue. This effect is visually demonstrated in Fig.~\ref{fig: periodogram_issues}.}
        
        \subsubsection{Uncertainty}
        \label{sec:uncertainty}
        In addition to the issue with template bias, we note that in many publications, period-uncertainty quantification is not treated robustly --- nor, in some cases, reported at all. This surprising omission is dismayingly common in both older studies which use the traditional method to calculate the period \citep[e.g.,][]{Bailey_1902,Wemple1932} and more recent studies with periodogram-based determinations \citep[e.g.][]{Siegel2015}. 
        A clear understanding and treatment of uncertainties is of great importance for modern studies, as long-term data can be used to study the evolution of stars \cite[e.g.,][]{Lee1991_HBevolution}. In the current state of affairs (with numerous measurements that lack uncertainty estimates), it is difficult to assess the quality of conclusions that rely on such historical data points. 
        Clear understanding of the uncertainty in period is also essential for studying other parameters that covary with the period in the fitting process or that are affected by a propagated uncertainty from the period. A demonstration of this issue and improvements that result from using \PIPS are presented in Sec.~\ref{sec: stellar properties}.

        Statistically speaking, the uncertainty in frequency space is often defined by the width of the periodogram peak, or more specifically, the half-width at half-maximum height (HWHM). This value is a function of the span of the data, and proper weight must be added considering the quality and size of the data according to \cite{Gregory2001_stats}:
        \begin{equation}
        \sigma_f \approx f_{1/2} \sqrt{\frac{2}{N(\text{SNR})^2}} .
        \end{equation}
        Hence,
        \begin{equation}
            \label{eq: stat_uncertainty}
            \sigma_P 
            = \left| \frac{-(1/f_\mathrm{peak})\sigma_f}{f_\mathrm{peak}}\right| 
            \approx \frac{f_{1/2}}{f_\mathrm{peak}^2}\sqrt{\frac{2}{N\cdot \mathrm{rms}[(y_j-\overline{y})/\sigma_j]^2}} \ .
        \end{equation}
        Although this formula is reasonable considering the effect of period folding,\footnote{The deviation due to the incorrect period is multiplied by $M$ if the dataset spans over $M$ pulsation cycles. One may thus assume the uncertainty should become linearly smaller as the baseline of the data becomes longer and $M$ becomes larger.} observations (e.g., Fig.~\ref{fig: periodogram_issues}) clearly show that this method does not provide an accurate view of the uncertainty.
        \cite{VanderPlas_2018} also suggests that this method should be avoided and introduces various methods to calculate false-alarm probability (FAP), with a conclusion that there is no universal method to directly estimate uncertainties from the periodogram.
        
        Instead, as the second step after the peak detection in a periodogram, one can perform a full-parameter optimisation. This process treats the period as a free parameter, rather than as a fixed value at the test location. The goal is not \revv{only} to determine the period, but to \revv{also} measure its uncertainty. 
        This full-parameter fitting can be achieved by either linear regression or more computationally expensive Bayesian methods, such as a Markov chain Monte Carlo (MCMC) exercise. For the linear regression method, one can estimate the size of the uncertainty from the covariance matrix $\Sigma_{ij}$ provided by linear regression methods as
        $\sigma_{P} \approx \sqrt{\Sigma_{i_P,i_P}}$.
        Similarly, the uncertainties of other parameters can be approximated by taking the square root of each element in $\Sigma_{ii}$, where $i$ denotes the index of each parameter.
        \rev{This full-parameter optimisation is commonly employed in the literature when the uncertainty values are reported \citep[e.g.,][]{Eyer2017_GaiaDR1_var}, but the computational cost becomes increasingly high as the data grow.}

        \subsubsection{Significance}
        \label{sec:SDE}
        The methods described in Sec.~\ref{sec:periodogram_concept} and Sec.~\ref{sec:uncertainty} to detect the period and estimate its uncertainty do not distinguish between physical signals and false positives in the noise. A method that provides a statistical measure of the proposed signal's reliability is required, and \PIPS thus provides signal-detection efficiency (SDE) as a statistic.
        
        \rev{What qualifies as a ``significant'' SDE is not generally derived empirically. For instance, a common significance threshold is the SDE below which  99\% of false-positive signals fall  (e.g., \citealt{Hippke_2019_TLS}). Hence, an SDE threshold is in principle algorithm-specific.}
        
        Following, for example, \cite{Kovacs_2002_BLS} and \cite{Hippke_2019_TLS}, SDE is evaluated to measure the relative strength of a signal with respect to the background noise level
        \begin{equation}
            \label{eq:SDE}
            \text{SDE}(P) = \frac{\text{SR}(P)-\langle\text{SR}(P)\rangle}{\sigma_\text{SR}}\ ,
        \end{equation}
        where SR (signal residual) is the periodogram power in Eq.~\ref{eq: periodogram} normalised to $1$ [i.e., $\text{SR}(P)=\mathcal{P}(P)/\mathcal{P}_\text{max}$]. The values of $\langle\text{SR}(P)\rangle$ and $\sigma_\text{SR}$ are the mean and standard deviation of the SR evaluated over the period search range $P_\text{min}<P<P_\text{max}$, respectively. As pointed out by \cite{Kovacs_2002_BLS}, finite evaluation range may bias the SDE, and it is therefore suggested that one use the SDE under the same set of parameters (e.g., the same period search range) and only use the SDE as a relative statistic for each data point in a dataset. A larger SDE value indicates a larger statistical significance, and a smaller value is often associated with \rev{noise, false positives for nonvarying objects, or astrophysical false positives fitted with an incorrect template.} 

\section{A new approach: likelihood periodogram}
\label{sec:likelihood}
        \begin{figure}
            \centering
            \includegraphics[width=\linewidth]{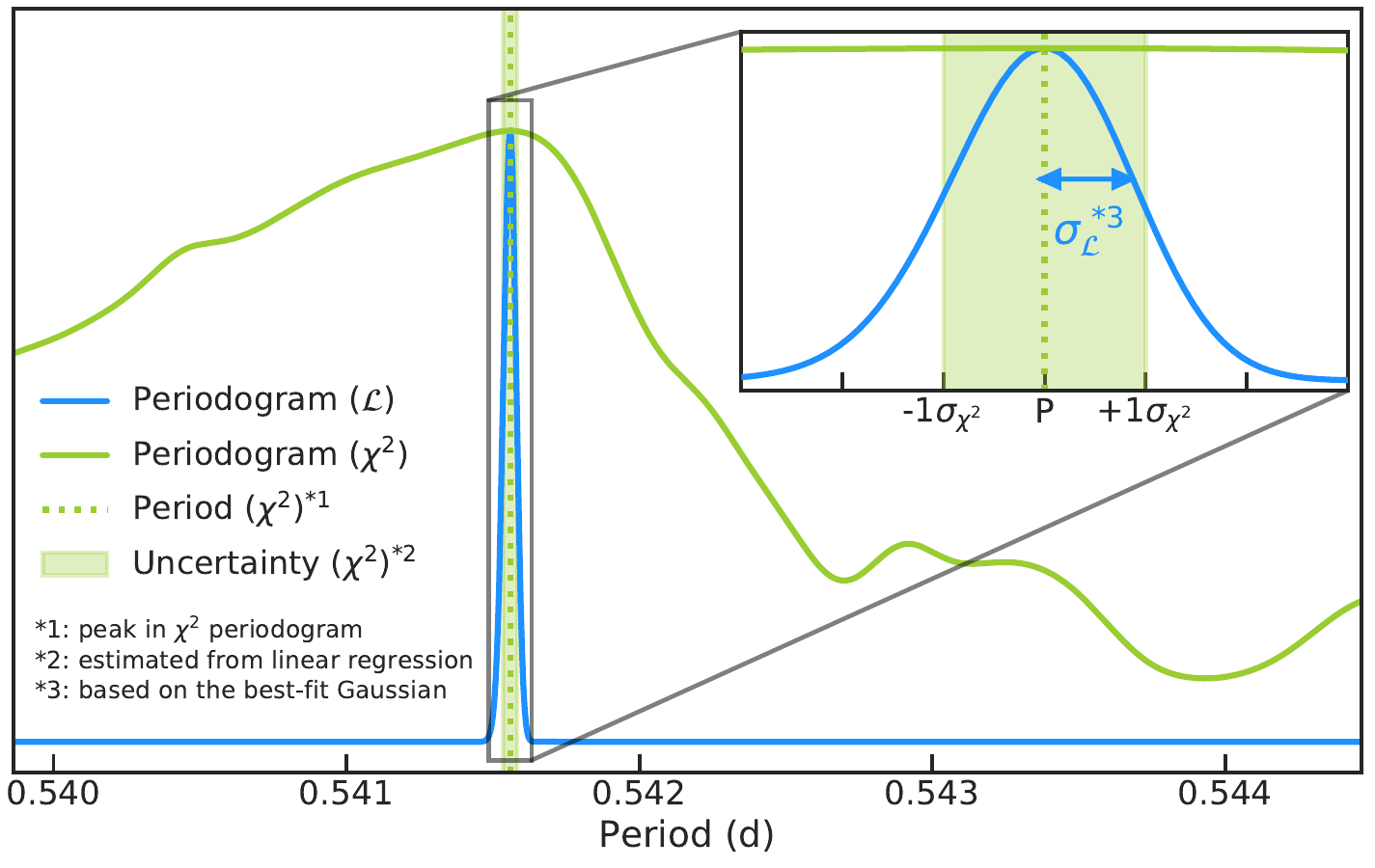}
            \caption{Likelihood periodogram (blue curve) and $\chi^2$ periodogram (light-green curve), normalised to equal height at the peak. Unlike the traditional $\chi^2$ periodogram, the likelihood periodogram can be directly interpreted as a probability density after normalisation, and the width of the peak accurately represents the uncertainty of the detected period.}
            \label{fig:likelihood}
        \end{figure}

        In Sec.~\ref{sec:uncertainty} and Sec.~\ref{sec:SDE}, we discussed important issues with the traditional $\chi^2$ representation of the periodogram, namely that its spectral density does not directly translate into scientifically important quantities such as signal amplitude, uncertainty, or statistical significance. 
        We discussed the existing solutions to these issues, such as more generalised or specialised templates, independent Monte-Carlo sampling in the light curve model's parameter space near the peak of the periodogram, and evaluation of Signal Detection Efficiency (SDE). These solutions are the current standard for scientific analysis, but have their limitations due to the computational cost and lack of statistical understandings.
        \rev{To address the existing issues and expand the possibility set of transient analyses in the era of large, data-rich all-sky surveys}, we propose a new approach, the \emph{Likelihood periodogram}. 

        Likelihood periodogram is a new representation of the $\chi^2$ periodogram, and is achieved by converting the $\chi^2$ values to Gaussian likelihoods for each data point, and then taking the product of the likelihoods across all data points. The likelihood is thus defined as
        \begin{eqnarray}
            \label{eq:gaussian-likelihood}
            \mathcal{L}(P_\text{test}\given x,y,\sigma_y) &=& \prod_i^{N_\text{data}}
            \frac{1}{\sigma_i\sqrt{2\pi}}\exp\left[-\frac{\left(y_i-y_{i,\text{fit}}\right)^2}{2\sigma_i^2}\right]\\
            &\propto& \exp\left(-\frac{1}{2}\sum_i\chi^2_i\right)\ .
        \end{eqnarray}
        This form, \rev{inspired from Bayesian statistics, comes with great benefits thanks to its own statistical properties. We discuss these benefits below, demonstrating how this new approach can address the issues with the current standard.}

        \subsection{Concept}
        \label{sec:likelihood_concept}
        \begin{figure}
            \centering
            \includegraphics[width=\linewidth]{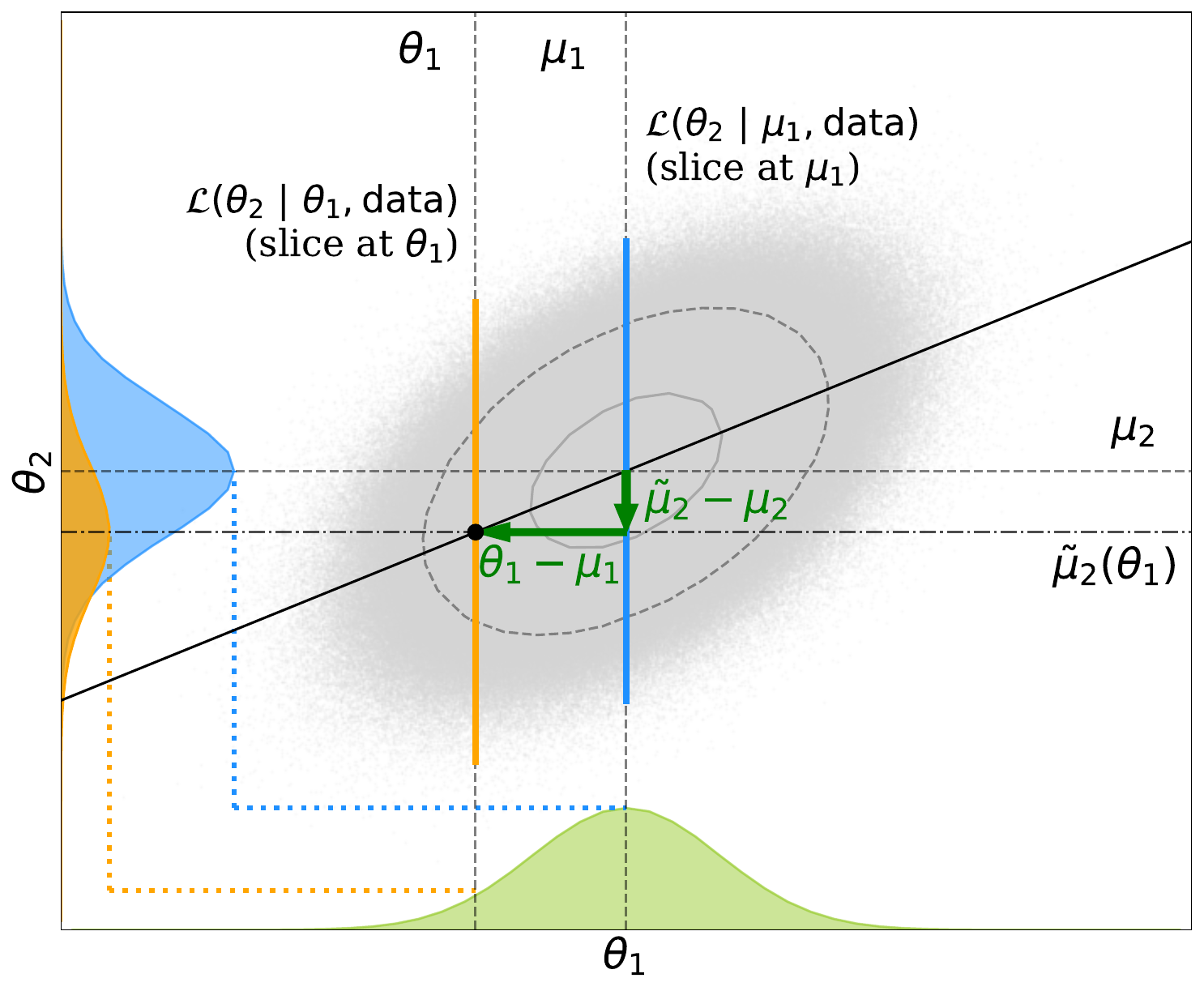}
            \caption{\rev{A visualisation of Eq.~\ref{eq:bivariate_slice}. The marginalised distribution of the parameter of interest $\theta_1$ (green) for the given 2D posterior samples (grey dots and contour) can be calculated without intense sampling if one can find the optimal value of $\theta_2$ ($\equiv \tilde\mu_2$, the centre of distribution for each slice) at each $\theta_1$. The height of normal distributions (blue and orange curves, centred at $\mu_2$ and $\tilde\mu_2$, respectively) is proportional to the marginalised likelihood for $\theta_1$.}}
            \label{fig:bivariate_marg}
        \end{figure}

        \rev{
        In model fitting --- which we often employ in studies of RR Lyrae stars, Cepheids, eclipsing binaries, or exoplanet transits --- data quality, model parameterisation, and robust sampling algorithms create near-Gaussian parameter posterior distributions. This is often assumed, and thus the uncertainty is commonly reported using one standard deviation ($1\sigma$) of the Gaussian posterior probability density function (PDF)\footnote{One can tell that this is a common practice by the fact that only one value of uncertainty is generally reported in the literature.}. 
        This is an important but often-overlooked fact in the period analysis of variable objects in astronomy; it will enable us to expand our understanding of the periodogram significantly.}

        \rev{For the multivariate posterior PDF, the likelihood that a parameter $\theta_i$ (e.g., period) is consistent with the observed data ($x$, $y$, and $\sigma_y$) is calculated (often from MCMC samples) by marginalising the posterior PDF over all other parameters. For the bivariate (two-parameter) case, for instance, we obtain the likelihood of $\theta_1$ by integrating over $\theta_2$:
        \begin{equation}
            \label{eq:bivariate_marginal}
            \mathcal{L}(\theta_1 \given x,y,\sigma_y ) = 
            \int \mathcal{L}(\theta_1,\theta_2 \given x,y,\sigma_y )\ 
            \mathrm d\theta_{2}\ .
        \end{equation}
        When the posterior PDF $\mathcal{L}$ is exactly or well approximated by a multivariate Gaussian, the likelihood of $\theta_2$ at fixed $\theta_1$ can be written as
        \begin{equation}
            \label{eq:bivariate_slice}
            \mathcal{L}(\theta_2\given \theta_1,x,y,\sigma_y) \propto \alpha(\theta_1)\exp\left[-\frac{(\theta_2 - \tilde\mu_2)^2}{2\beta^2}\right] ,
        \end{equation}
        where the local mean $\tilde\mu_2$ is the best-fit value of $\theta_2$ at fixed $\theta_1$ (see Fig.~\ref{fig:bivariate_marg}).
        The height of the Gaussian $\alpha$ is a function of $\theta_1$, and $\beta$, the standard deviation of $\theta_2$ PDF sliced at fixed $\theta_1$ is independent from $\theta_1$ or $\theta_2$. A more detailed derivation of Eq.~\ref{eq:bivariate_slice} is provided in Appendix~\ref{sec:appendix_gaussian}.
        }

        \rev{
        In other words, the shape and the relative size of the marginalised likelihood $\mathcal{L}(\theta_1\given x,y,\sigma_y)$ can be obtained by the value of $\alpha$ (arbitrarily normalised likelihood) at $(\theta_1,\theta_2)=(\theta_1,\theta_{2,\mathrm{fit}})$. This feature is beneficial because the value of $\theta_{2,\mathrm{fit}}$ can be obtained by the maximum-likelihood fitting in the $\theta_2$ space at each $\theta_1$. Generalising this concept to the $N$-parameter case, we obtain
        \begin{equation}
            \label{eq:general_likelihood}
            \mathcal{L}(\theta_1\given x,y,\sigma_y) \propto
            \mathcal{L}\left(x,y,\sigma_y \given \theta_1, \theta_{2,\mathrm{fit}},\theta_{3,\mathrm{fit}},\cdots\right)\ ,
        \end{equation}
        and therefore
        \begin{equation}
            \label{eq:general_loglik}
            \log\mathcal{L}(\theta_1\given x,y,\sigma_y) =  -\frac{1}{2}\sum_i \chi_i^2\ + \gamma\,
        \end{equation}
        with a constant offset $\gamma$.
        }

        \rev{
        Our formulation suggests that \emph{the periodogram provides all the statistical tools} we need to assess the posterior PDF. This added statistical value in the periodogram now allows us to address the existing issues directly in more informed and far more cost-effective ways.
        }
        
        \subsection{Uncertainty}
        In Fig.~\ref{fig:likelihood}, we present an example of the likelihood periodogram, normalised to the same height as the traditional $\chi^2$ periodogram. 
        From the comparison between the estimated uncertainty (light-green fill) and the size of the standard deviation in the likelihood periodogram's Gaussian shape (blue arrows), it is evident that the likelihood periodogram accurately portrays the (relative) statistical significance.

        \rev{This example and the equations in Sec.~\ref{sec:likelihood_concept} suggest}, more importantly, that the statistical significance of the peak can be directly calculated from the likelihood periodogram. This is a strong advantage of the likelihood periodogram over the $\chi^2$ periodogram. Computing the likelihood periodogram is, therefore, highly impactful in a time-series analysis, especially when one aims to detect a new candidate for previously unknown variable stars.

        The significantly narrower (and hence more accurate) peak causes a problem with computation. As we discussed in Sec.~\ref{sec:peak_selection}, the sampling width has to be smaller than the width of the peak, and a significantly narrower peak requires a large number of calculations. We address this issue by \rev{carefully spacing our samples in the periodogram as described in Sec.~\ref{sec:peak_selection}}. When the likelihood periodogram is enabled in \PIPS, it (by default) uses the period and uncertainty detected using the original $\chi^2$ periodogram to pinpoint the location to sample the likelihood periodogram. \PIPS determines the sampling resolution and range using the fact that the size of the estimated uncertainty is comparable to the half-width of the peak.

        \begin{figure}
            \centering
            \includegraphics[width=\linewidth]{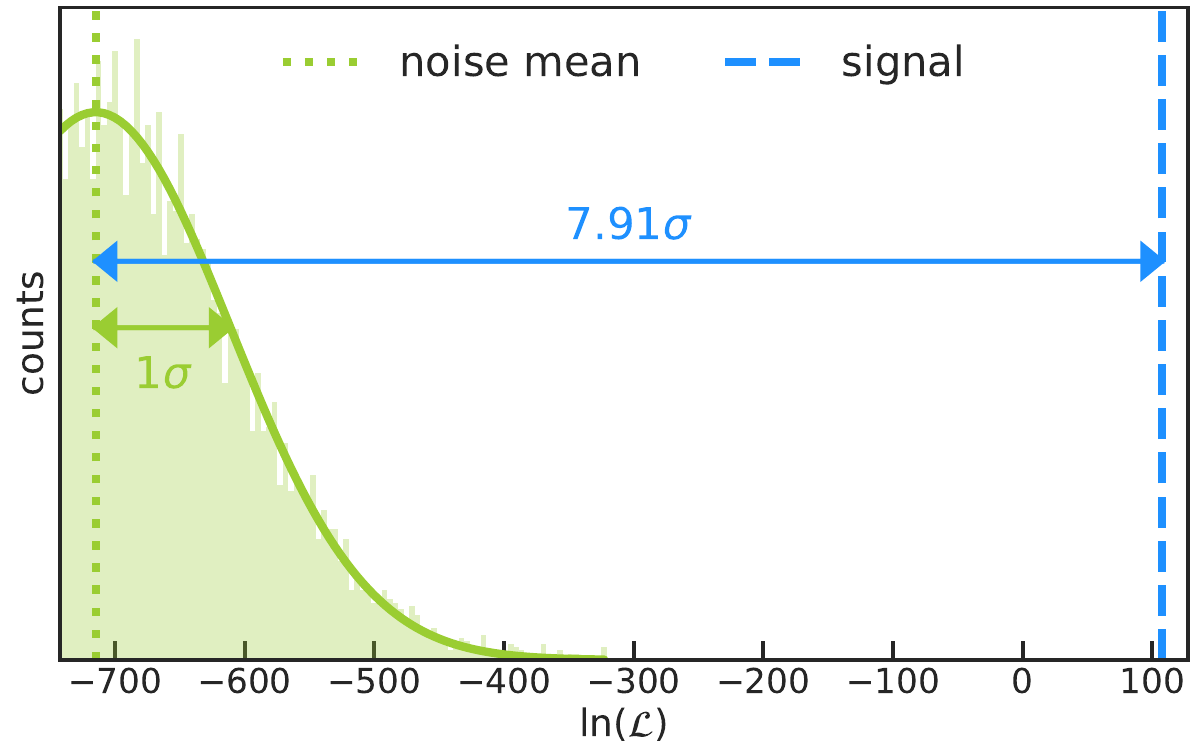}
            \caption{An example significance test in which the log-likelihood value of the detected signal (blue dashed line) is compared to the background noise (light green). The noise values are simulated by shuffling the data and measuring the log-likelihood at a random location (i.e., bootstrapping). The separation of the signal from mean noise is significantly (7.9 times) larger than the dispersion of the noise, indicating that the detected signal is statistically significant.}
            \label{fig:likelihood-sig}
        \end{figure}

        \subsection{Statistical Significance}
        Once the likelihood periodogram is calculated, the most probable period and its uncertainty can be directly calculated by fitting a Gaussian function to the periodogram. A statistical test can then be performed to quantify the significance of the detected peak. For instance, a likelihood-ratio (LR) test \citep[e.g.,][]{Buse1982_LikelihoodRatio} can be performed, with the LR defined as 
        \begin{equation}
            \label{eq:likelihood-ratio}
            \mathrm{LR}_\text{peak} = \ln\left[\frac{\mathcal{L}(P_\text{peak})}{\langle\mathcal{L}_\text{noise}(P)\rangle}\right] = 
            \ln\left[\mathcal{L}(P_\text{peak})\right] - \ln\left[\langle\mathcal{L}_\text{noise}(P)\rangle\right],
        \end{equation}
        where $\mathcal{L}(P_\text{peak})$ is the likelihood value at the peak, and $\langle\mathcal{L}_\text{noise}(P)\rangle$ is the mean of likelihood values at the background (i.e., noise). Calculating the noise likelihood can be accomplished by bootstrapping: shuffling the magnitudes while keeping the uncertainty values attached to each data point and randomly sampling the likelihood values across the period search-range serves as a simulation of the background noise. Comparison of this $\text{LR}_\text{peak}$ value to the noise level $\sigma_\text{LR, noise}$ quantifies the significance \rev{of the detected signal}. The statistical significance, represented by the $Z$-score, is thus
        \begin{equation}
            Z_\text{peak} = \frac{\text{LR}_\text{peak}}{\sigma_\text{LR,noise}} \ .
        \end{equation}
        Fig.~\ref{fig:likelihood-sig} shows the significance analysis for the peak value in Fig.~\ref{fig:likelihood}. The $Z$-score is 7.91 (i.e., a $7.91\sigma$ detection), or $p<0.00001$,\footnote{Here we are referring to the statistical $p$-value, not to be confused with the exponent in the super-Gaussian model in Eq.~\ref{eq:supergaussian}.}, and the detected peak is said to be statistically significant.

\section{PIPS: algorithm and implementation}
    \label{sec: algorithm}
    \rev{
    We implement the above methods for period detection in \PIPS\  --- an advanced, open-source \texttt{Python} platform for the period analysis of astronomical data.
    \PIPS provides an all-in-one workflow for automated period detection, which we rigorously test in the next section. In addition, \PIPS gives highly optimisable and customisable tools for advanced users, thereby enabling tightly-controlled analyses for specific needs.
    In this section, we discuss the methods for automated period detection implemented in \PIPS, a number of light-curve templates that we implement by default, and additional features related to period detection. A selection of additional features for further analysis, such as stellar-parameter estimations or long-term period-change analysis, are discussed in later sections.
    }

    \subsection{Peak Selection and Refinement}
        \label{sec:peak_selection}
        Detecting the period of time-series data requires a robust algorithm to select a peak in the periodogram and its location as precisely as possible. As commonly known in many optimisation problems \citep[for review, see ][]{kochenderfer2019algorithms}, sampling size in $\chi^2$ space can have a significant impact on the results --- sampling that is too coarse will result in a missed peak spike, while increasing the sampling size directly impacts the computational cost. Our peak-selection algorithm provides a good compromise between these two, successfully detecting the period for most cases without excessively increasing the computing cost. 
        
        \PIPS initially generates the periodogram\footnote{Fitting of light-curve models is performed by a manually implemented linear problem solver or by linear regression with the \texttt{curve\_fit} function in \texttt{SciPy} \citep{2020SciPy-NMeth}.} within a specified period-search range, which is sampled densely enough so that the spacing of the sampling grid is smaller than the expected peak width. The expected peak width is calculated following the discussion by \cite{VanderPlas_2018}.
        Peak size in frequency space can be estimated using the baseline length $T$: $\Delta f = 1/T$. Defining both edges of the width at peak $f$ by $f_1 = f - \Delta f/2$ and $f_2 = f + \Delta f/2$, the conversion between period and frequency $p = 1/f$ gives the expected width size in period space,
        \begin{equation}
            \label{eq:peak_width_derivation}
            \Delta p = p_1 - p_2 = \frac{1}{f_1} - \frac{1}{f_2} = \frac{f_2 - f_1}{f_1 f_2} = \frac{\Delta f}{f^2 - \left(\Delta f\right)^2/4}\ .
        \end{equation}
        Combining this result with $\Delta f = 1/T$, we obtain 
        \begin{equation}
            \label{eq:peak_width}
            \Delta p = \frac{1/T}{(1/p)^2 - (1/4T^2)} = \frac{p^2T}{T^2-p^2/4}\ .
        \end{equation}
        Oversampling with a grid size smaller than this peak width \rev{--- by a factor of $N_0$ ---} will ensure that the peak is detected, so \PIPS uses this as the grid size,\footnote{We formulate this grid size in period space because it is less sensitive to the numerical error with finite grid size. In most cases, this can be approximated by adopting an equally spaced grid in frequency space and taking the reciprocal, but for long-period objects (i.e., $f \ll 1$), this can cause a significant numerical error, and caution is needed.}
        \begin{equation}
            \label{eq:grid_size}
            (\Delta p)_\text{grid} = \frac{\Delta p}{N_0} = \frac{p^2T}{N_0(T^2-p^2/4)}\ ,
        \end{equation}
        and the resolution of periodogram $R_\text{all}$ within the specified search range between $p_\text{min}$ and $p_\text{max}$ becomes
        \begin{equation}
            \label{eq:sample_size}
            R_\text{all} = \frac{p_\text{max}-p_\text{min}}{(\Delta p)_\text{grid}} = \frac{N_0\left(p_\text{max}-p_\text{min}\right)\left(T^2-p^2/4\right)}{p^2_\text{min} T}\ .
        \end{equation}
        Once the periodogram with resolution $R_\text{all}$ is generated, the top $N_\text{peak}$ independent peaks are selected from the initial periodogram. The independence of these peaks is checked by ensuring a distance of at least $(\Delta p)_\text{min}$ between peaks.
        The periodogram only around those peaks is then densely sampled (with a sample resolution $R_\text{peak} \gg N_0$ within $\Delta p_\text{peak}$ at each peak). Usually this ensures a fine sampling around the true peak where $(\Delta p)_\text{grid} \ll \sigma_p$ (for the estimation of $\sigma_p$, see Sec.~\ref{sec:uncertainty}), and when this condition is not satisfied, \PIPS raises a warning to suggest an increased peak resolution, $R_\text{peak}$.

    \subsection{\rev{Light-Curve} Models}
        \label{sec:models}
        \rev{We designed} \PIPS \rev{as} a general platform for period detection with customisable models. That is, while we implement a few fundamental generative functions, \emph{any} template for light curves can be used. For instance, for objects whose light curves have well-known analytic forms, using the true physical model will minimise template bias \rev{ --- assuming that any long-term instrumental trends and outliers have already been removed.\footnote{Outliers can be removed by user-defined criteria using a function in \PIPS. Future versions of PIPS will include long-term detrending with \texttt{lightkurve}: \url{https://docs.lightkurve.org/}}} Details for using custom functions can be found in the online documentation.\footnote{\url{https://pips.readthedocs.io/en/latest/}} In this section, we describe the three functions we choose as pre-implemented models for general-purpose use.
        
        \subsubsection{Fourier series}
        \label{sec:model_Fourier}
        By default and following the convention in the literature \citep[e.g.,][]{GaiaDR2_variability}, \PIPS uses a multiterm Fourier series of the form
        \begin{equation}
            \label{eq: Fourier fitting}
            y_\mathrm{fit}(x; P_\mathrm{test}) = A_0 + \sum_{k=i}^{K_\mathrm{max}} \left[a_k \cos\left(\frac{2\pi k}{P_\mathrm{test}}x\right) 
                        + b_k \sin\left(\frac{2\pi k}{P_\mathrm{test}}x\right)\right] \ ,
        \end{equation}
        to approximate light curves. Here, $y_\text{fit}$ is the fitted magnitude at given time $x$. The constant offset $A_0$ makes the fit independent of the distance modulus and/or the luminosity of the object under study, and the coefficients $a_k$ and $b_k$ control the amplitude of each $k^{\rm th}$ harmonic. These free parameters (i.e., $A_0, a_k,$ and $b_k$) are optimised so that the phase-folded light curve at each test period $P_\text{test}$ best matches the observed magnitudes. The series length, $K_\text{max}$, controls the number of harmonics in the light curve: a larger value of $K_\text{max}$ results in a better approximation to sharper peaks and smaller features in a light curve, while making the fit more sensitive to the noise and possibly lowering the signal detection rate. 
        In our first use of this algorithm with RR Lyrae stars by \cite{H&M2020}, we set $K_\mathrm{max}=5$, based on a cross-validation test using {\it Gaia} data.
        
        \subsubsection{Super-Gaussian} 
        \label{sec:model_SG}
        \begin{figure}
            \centering
            \includegraphics[width=\linewidth]{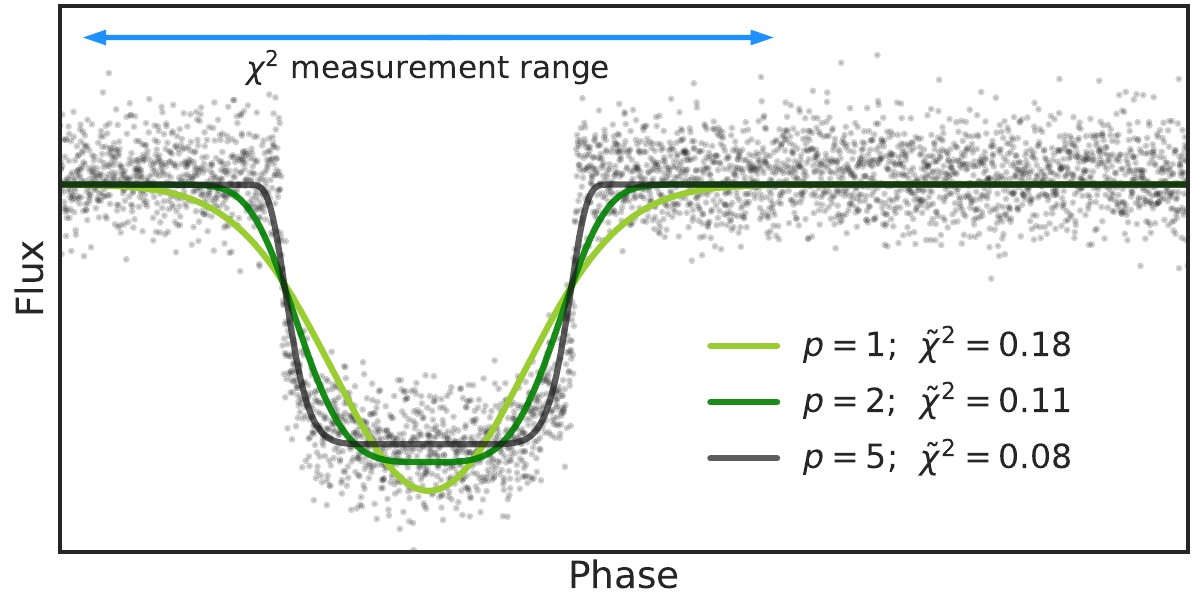}
            \caption{Best-fit super-Gaussian (SG) curves to artificially generated exoplanet transit data (grey dots). SG with $p=5$ produces a better fit to the data than $p=1$ (i.e., Gaussian). The $\chi^2$ values are measured only near the transit phase (indicated by an arrow) to illustrate the difference between different choices of $p$.}
            \label{fig:SG_demo}
        \end{figure}
        
        For some types of periodically varying objects, such as transiting exoplanets or binary stars, the light curve can have a narrow, sharp feature, which cannot be well approximated by a Fourier series without massively increasing the number of free parameters (i.e., setting $K_\text{max}$ to a large value). To account for these types of features, we experimentally implement the Super-Gaussian model (SGM), which we formulate as
        \begin{equation}
            \label{eq:supergaussian}
            y_\text{fit}(x;P_\text{test}) = B_0 - B_1  \exp\left[-\left(\frac{(\phi-\mu)^2}{2\sigma^2}\right)^p\right] \ .
        \end{equation}
        In Eq.~\ref{eq:supergaussian}, $\phi = (x \mod P_\text{test})$ is the phase-folded time, $B_0$ is the mean of the nonvarying (i.e., flat) part of the light curve, and three parameters ($B_1$, $\mu$, and $\sigma$) respectively determine the height, location, and width of the feature. The power parameter, $p$, forces the original Gaussian function to a more square-like shape. 
        The effect of changing the $p$-value is demonstrated in Fig.~\ref{fig:SG_demo}, in which the normalised $\chi^2$ ($\tilde{\chi^2}=\chi^2/\chi^2_0$) to the best-fit curve around the transit well is calculated.
        
        This model can be converted to a log-linear form, which is computationally less expensive,
        \begin{align}
            \log{\left(y_\text{fit}-B_0\right)} &= \log(-B_1) - \frac{1}{(2\sigma^2)^{p}}\left(\phi-\mu\right)^{2p} \nonumber \\
            &= \theta_0 + \theta_1 \phi + \theta_2 \phi^2 + \theta_3\phi^3 + \cdots + \theta_{2p}\phi^{2p}\ .  \label{eq:SG-loglinear}
        \end{align}
        This form has a computational advantage because the ``exact" solution can be calculated as a linear-algebra problem.
        It should be noted, however, that using linear algebra to find the solution \emph{does not} guarantee that the $\chi^2$ value is minimised \citep[for a review, see, e.g.,][]{AL-Nahhal2019_Gaussian}. Moreover, this generalised form in Eq.~\ref{eq:SG-loglinear} is only equivalent to the original form in Eq.~\ref{eq:supergaussian} when all free parameters ($\theta_0, \theta_1, \cdots$) satisfy the corresponding function of the original parameters $B_1$, $\mu$, and $\sigma$. This means that the log-linear form in Eq.~\ref{eq:SG-loglinear} produces a higher level of noise with less-sharp peaks, even though it is computationally much more efficient to produce the periodogram. 
        Because of this issue, we use nonlinear, iterative, least-squares fitting by default.
        
        \subsubsection{Gaussian Mixture} 
        \label{sec:model_GMM}
        We also implement another well-known generative function: the Gaussian Mixture model (GMM), which can take multiple narrow independent peaks (or dips). As a result, we expect this model to work better than Fourier models when the input light curve has a combination of wide and narrow features. 
        The GMM can be expressed as
        \begin{equation}
            \label{eq:GMM}
            y_\text{fit}(x;P_\text{test}) = C_0 + \sum_{k=1}^{K_\text{max}}c_k \exp{\left(-\frac{(\phi-\mu_k)^2}{2\sigma_k^2}\right)}\ ,
        \end{equation}
        where the constant offset ($C_0$), amplitude ($c_k$), location ($\mu_k$), and width ($\sigma_k$) are fitted as free parameters for each $k^{\rm th}$ term up to the specified number of terms, $K_\text{max}$.
        For convenience, we combine the SGM and GMM, and implement the Super-Gaussian Mixture model in \PIPS, as
        \begin{equation}
            \label{eq:SGMM}
            y_\text{fit}(x;P_\text{test}) = C_0 + \sum_{k=1}^{K_\text{max}}c_k \exp{\left[-\left(\frac{(\phi-\mu_k)^2}{2\sigma_k^2}\right)^p\right]}\ .
        \end{equation}
        Setting $p=1$ reverts Eq.~\ref{eq:SGMM} back to the GMM (Eq.~\ref{eq:GMM}), while taking $K_\text{max}=1$ reverts it back to the SGM (Eq.~\ref{eq:supergaussian}).

        \subsubsection{The impact of model choice}
        \label{sec:model choice}
        As shown in Fig. \ref{fig: periodogram_issues} and Fig. \ref{fig:SG_demo}, light curves with discrepant morphologies are best fit by models that are appropriate for their specific morphologies. While a basic, two-term Fourier model would be able to provide a period estimate for any periodic data, the signal accuracy and strength will be severely diminished for a pathological choice of light-curve morphology (e.g., a very shallow exoplanet transit). Increasing the number of Fourier terms will likely provide a better fit to the data, but at the expense of computation time and an increased risk of overfitting. Hence, it is crucial to identify the optimal model choice for light-curve fitting; working with a subpar model will only provide incremental fitting improvement.
        
        \begin{figure}
            \centering
            \includegraphics[width=\linewidth]{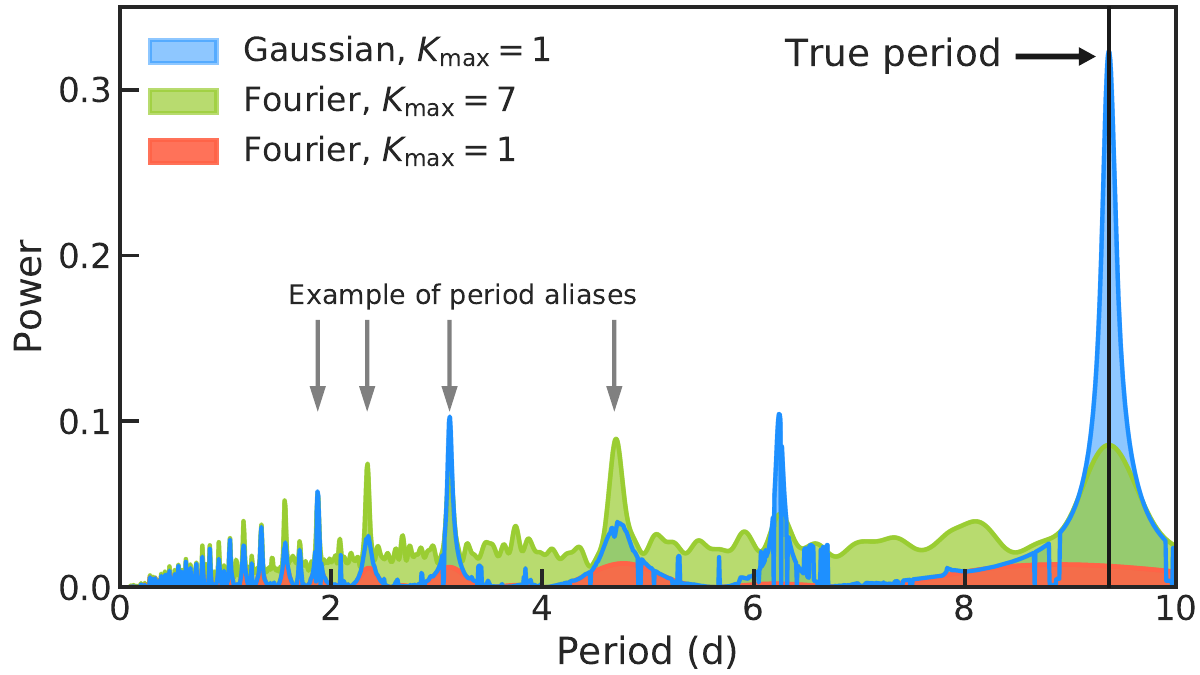}
            \caption{Benchmarking the effectiveness of different \PIPS models in determining the period of a transiting exoplanet. The single-term Gaussian model outperforms the Fourier model both in better constraining the true period and in deweighting the statistical importance of its harmonics.}
            \label{fig:model_comparison}
        \end{figure}
        
        This intuition is made more evident in Fig. \ref{fig:model_comparison}. Here, the advantage in choosing better models for a given dataset can clearly be seen --- a narrow periodogram peak near the true period, in addition to weaker aliasing at harmonics of the true period.

    \subsection{Multiperiodic Analysis}
        Some objects, such as double-mode pulsators or transiting binaries with one or both being a variable star on its own, exhibit a light curve that is a linear combination of two or more variations at different frequencies. Analysis of such objects requires decomposing each variation without losing the light-curve shape, and \PIPS achieves this by iteratively performing period analysis and removing the dominant component at each detected period \citep[prewhitening; e.g.,][]{Blomme2011_prewhitening}.
        
        The result of this analysis can be represented as an amplitude spectrum. As described in \cite{H&M2020}, this amplitude spectrum is \rev{a representation of consecutive prewhitening and} an improved and generalised version of the nonlinear discrete Fourier transform (DFT) method,
        \begin{equation}
            \label{eq:amplitude_spectrum}
            y = A_0 + \sum_i^{i_\text{max}} A_i\mathcal{F}_i\left(x,P_i;\{\theta_j\}_i\right)\ ,
        \end{equation}
        where $\mathcal{F}_i$ is the normalised model function (see Sec.~\ref{sec:models}) and $\{\theta_j\}_i$ is a set of best-fit model parameters at each period.
        The period $P_i$ at which the function is evaluated is determined by our main period-search method described in Sec.~\ref{sec:peak_selection}. This method is effective at detecting the pulsation modes in the resulting amplitude spectrum, since the harmonic components of the target frequencies at each pulsation mode are fitted (i.e., removed) at each iteration. 
        
        Note that $F_i$ has a subscript because the model can be changed for each period. For instance, one could search for the transiting exoplanet around a single-mode variable star by specifying $F_0=\text{Fourier}$ and $F_1=\text{SGM}$. Since each model has a unique behaviour that enables more optimised search for specific targets, this should greatly enhance the detecting capability of scientifically impactful objects beyond what general Fourier DFT allows us to do.

\section{Period-Detection Validation}
    \label{sec: Validation}
    As described in Sec.~\ref{sec: algorithm}, \PIPS provides an automated period-detection algorithm. This algorithm is developed with the analyses conducted by large surveys in mind, and therefore understanding the performance of \PIPS as a function of various data quality and input parameters is essential. 
    In the following sections, we describe our approach and results for validating the performance of \PIPS. These tests are focused on validating the period and uncertainty determination capabilities of \PIPS, and we discuss best practices and optimal data structures as indicated by our results.

\subsection{Data Construction}
    \label{sec: data construction}
    In order to test the accuracy of \PIPS for data of varying quality, cadence, and quantity, we generate artificial data based on samples from OGLE \citep{Soszynski_2009_OGLE_LMC}. These samples provide highly complete and dense phase-folded light curves, which allow us to generate artificial data with realistic light-curve shapes. For the OGLE samples, we take the top 20 results (16 RRab and 4 RRc stars) with the lowest reported period uncertainty.
    
    We generate a high-cadence set of observations for a single period by phase-folding the original OGLE data using the period provided by OGLE. We then extrapolate to generate a high-cadence set of observations over the desired period of time\footnote{\revv{This extrapolation allows us to generate the artificial data without analytic functions. The method provides sufficient data for this work since our tests mainly focus on poorly sampled, noisy data. One can also generate artificial data by using high-order Fourier series fitted to datasets from much more precisely calibrated surveys, such as {\it Kepler} \citep{Kepler_RRL_2010} when more physically ``true" mock data are desired.}}.
    To create an artificial dataset for a specified number of observations ($N$), days between observations (DBO), and SNR, the extrapolated dataset is sampled at the desired observation times.
    Gaussian noise of the desired size is then added to each observation's magnitude, and we define the uncertainty of the constructed data by adding the original uncertainty and the noise in quadrature. 
    The size of the resulting uncertainty is used to measure the SNR.
    We define a parameter, DBO-to-Window Ratio (DWR), as the ratio of the average number of days between observations (DBO), and the length of the observation window (W) for a single observation within which the observation's target time was randomly chosen. For instance, if a star is visible for 6\,hr ($= 1/4$ d)during the night and an observation was taken at a random time during that window each day, then DBO $ = 1$, W $= 1/4$, and DWR $ = 4$. This is used to determine when to sample the artificial data points. Target observation times for the new dataset are then generated, which follow the desired values of $N$, DBO, DWR, and SNR. 
    
    The new data are then analysed by \PIPS, and the period determined by \PIPS is compared to the one reported by OGLE. It is worth noting that, although the period reported by OGLE is only assumed to be the true period for the purpose of phase folding, it is the ``true'' period for the data given to \PIPS, since it is the period used to extrapolate the data during generation. 
    Deviation in the assumed period results in decreased SNR, but the period of the new set, and hence the difference between that period and the one calculated by \PIPS, is exact.

\subsection{\rev{Evaluation Metrics}} 
\begin{figure}
        \centering
        \includegraphics[width=\linewidth]{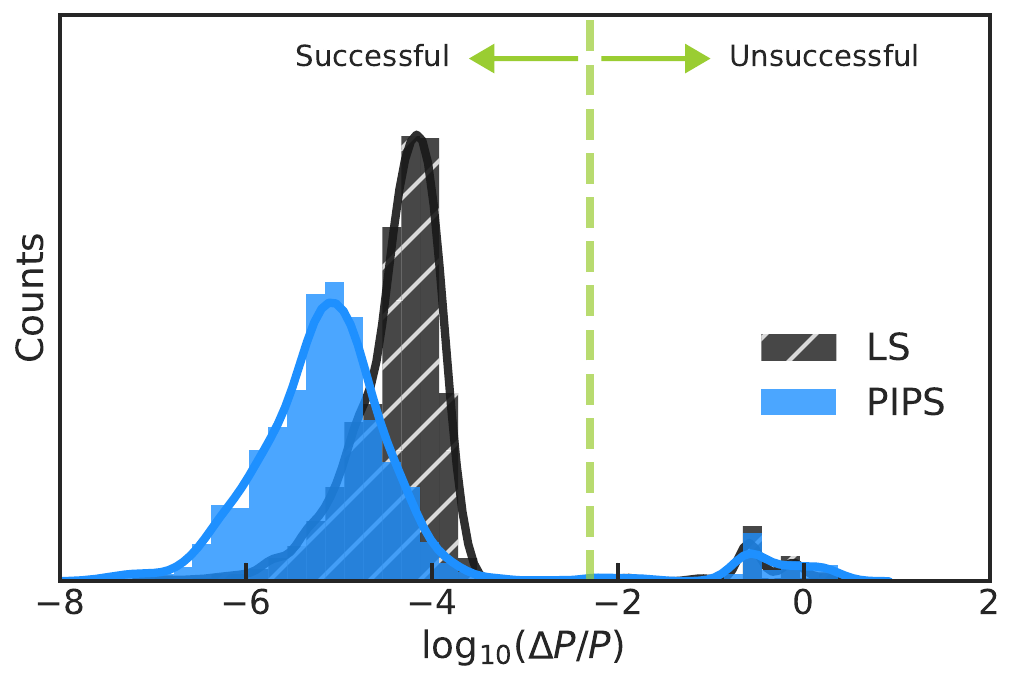}
        \caption{The same plot as Fig. 4 or \protect\cite{H&M2020} but in the relative deviation. The abscissa is in log base 10. Both LS and \PIPS result in a gap between ``successful'' and ``unsuccessful'' estimates. A more detailed distribution of the two populations in various parameter spaces is visualised in Fig.~\ref{fig:separation_plot}.}
        \label{fig: PIPS_LS_gap}
    \end{figure}

    We evaluate the performance of \PIPS using two metrics.
    \begin{itemize}
        \item \textbf{Success rate} -- the proportion of trials where \PIPS achieves a measurement within 0.5\%\footnote{We choose this threshold based on a clear separation as seen in Fig.~\ref{fig:separation_plot}.} of the true period
        \item \textbf{Relative deviation} --  the deviation relative to the accepted period for successful tests.
    \end{itemize}
    The threshold value of 0.5\% deviation from the true period is chosen to separate our results owing to the bimodal nature of the estimate's deviation from accepted values, as seen in Fig.~\ref{fig: PIPS_LS_gap} (and also in Fig.~\ref{fig:separation_plot}).

\subsection{PIPS Behaviour by Observation Parameters}
    \label{sec:test_obs_params}

    \begin{figure*}
        \centering
        \includegraphics[width=\linewidth]{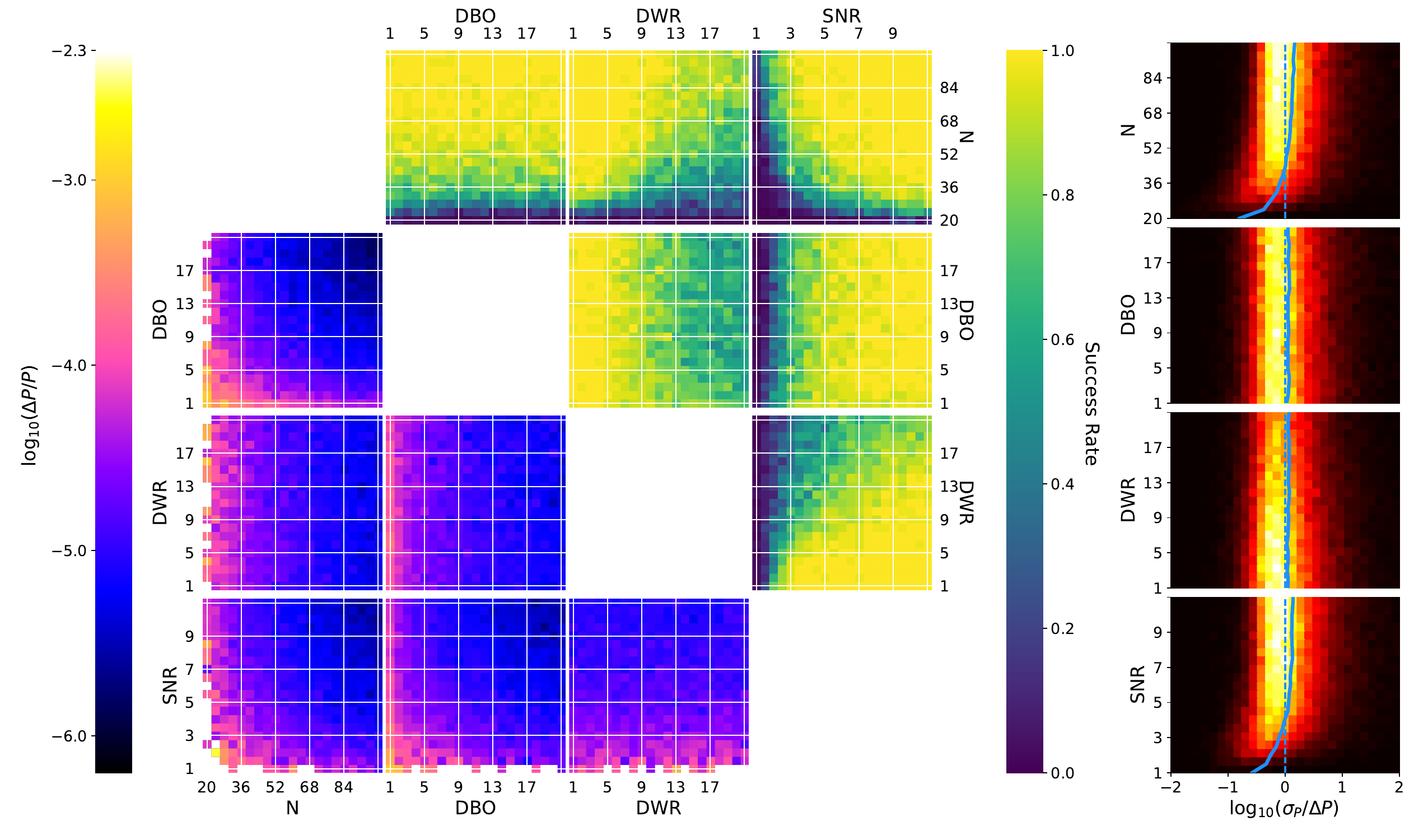}
        \caption{Left: the success rate and relative deviation from the known period using \PIPS on generated data. When not specified, $N = 50$, DBO $ = 7$, SNR $ = 5$, and DWR $ = 8$. Relative deviation values are calculated only for the ``successful" population. The empty pixels indicate that the success rate is 0, so the relative deviation cannot be determined. Right: Two-dimensional (2D) histograms of the ratio between estimated uncertainty and the deviation, with a $\log_{10}$ scale. The parameter ranges on the vertical axes are the same as in the left panel, and the data only include the ``successful" population.}
        \label{fig: All Paramaters}
    \end{figure*} 
    
    In our investigation, we identified four key attributes ($N$, DBO, DWR, and SNR) that can influence the results of a period detection, be it derived via LS, \PIPS, or otherwise. Here, we discuss how these attributes influence the results of \PIPS. 
    Our data-construction methods for a given set of attributes to be tested are outlined in Sec.~\ref{sec: data construction}.
    
    \paragraph*{$N$} ---
        Fig.~\ref{fig: All Paramaters} shows that increasing the number of observations increases both the success rate and the deviation size. This is expected, since the larger number of degrees of freedom in fitting should result in more accurate results.
        
    \paragraph*{DBO} ---
        Fig.~\ref{fig: All Paramaters} shows that this has no effect on the success rate but does decrease the deviation size. This may be due to increasing the total time over which observations are taken, thereby magnifying small errors in the period which in turn allows \PIPS to further constrain its results.
        Note that, while $N$ also increases the total time, it has an additional effect on the deviation size. For instance, in Fig.~\ref{fig: All Paramaters} $N = 20$ and DBO = 4 will cover the same total time as $N = 80$ and DBO = 1, but the deviation size at the latter point is less than the deviation size at the former.
        
    \paragraph*{DWR} ---
        For the tests run in this section, target times are determined using the number of observations and the days between observations. A random value between $-W/2$ and $W/2$ is added to each target time, where $W$ can be interpreted as the size of the window for a single observation.
        For a given DBO, DWR is determined by
        \begin{equation}
            \label{eq: test_equation}
            \mathrm{DWR} = \mathrm{DBO}/W\ .
        \end{equation}
        As DWR increases, the cadence becomes more regular, with observations becoming equally spaced as DWR approaches infinity. Fig.~\ref{fig: All Paramaters} shows that DWR has a negative correlation with success rate, and no relation with deviation size.
        The effects of low DWR can also be achieved by varying the time between observations using a nonrandom method, such as by advancing the time of observation by a set amount on consecutive nights, looping when necessary. \rev{When regularity of observation causes issues with aliasing, employing window functions may reduce the effect. We discuss the usage of window functions in the documentation.} 
    
    \paragraph*{SNR} ---
        Higher SNR increases the success rate and the deviation size. While it is difficult to control for the SNR when collecting data, it is possible to overcome an SNR of as low as 2 by focusing on other parameters, such as DWR, as seen in Fig.~\ref{fig: All Paramaters}.
            
\subsection{\PIPS Behaviour by Internal Parameters}
    \label{sec:test_PIPS_params}
    \begin{figure*}
        \centering
        \includegraphics[width=\linewidth]{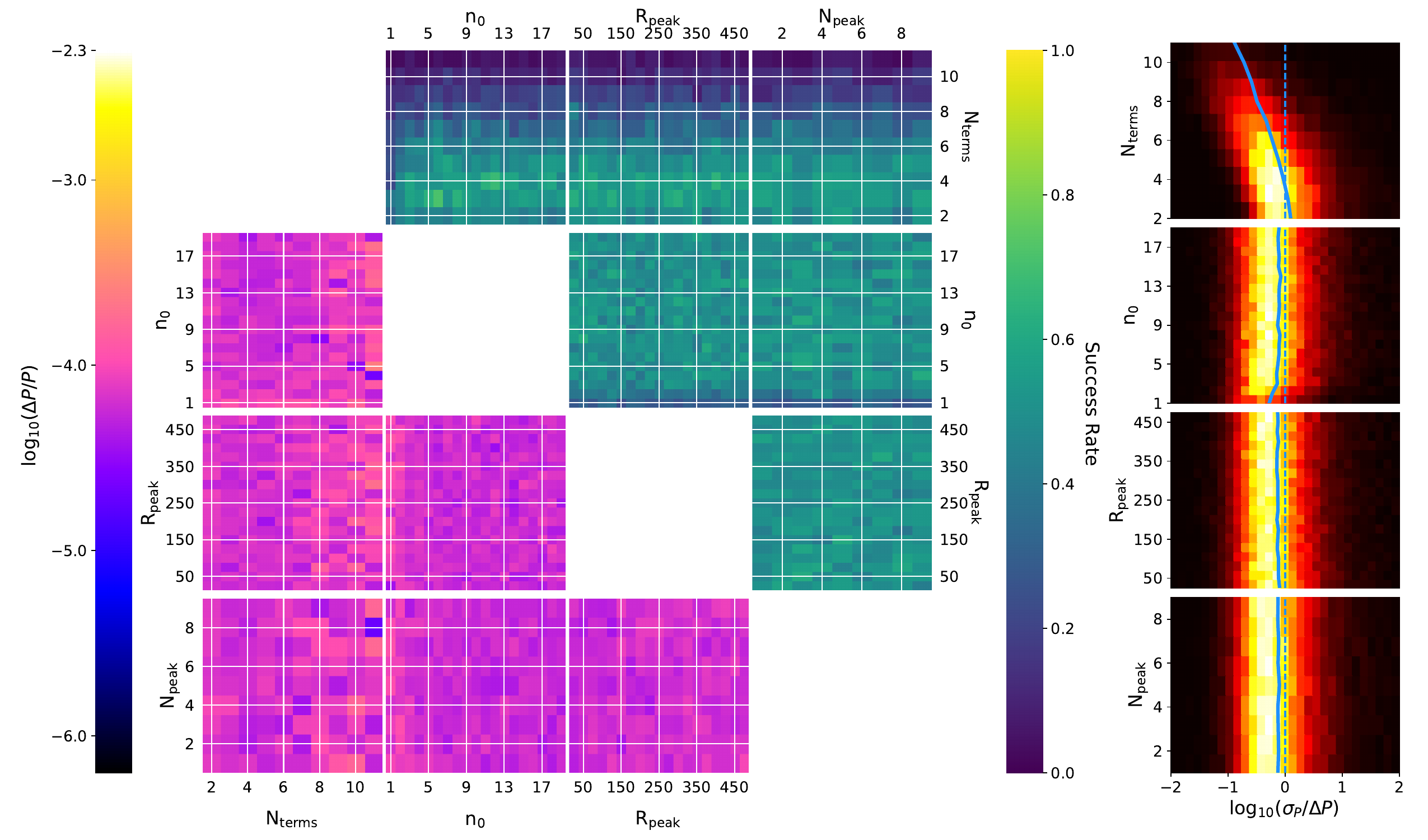}
        \caption{Left: The success rate and relative deviation from the known period using \PIPS on generated data. When not specified, $N_\text{terms}=5$, $N_\text{peak} = 7$, $R_\text{peak}=500$, $N_0 = 5$. The generated data have values of $N=35$, DBO = 7, DWR = 8, and SNR = 3, the effects of which are discussed in Sec.~\ref{sec:test_obs_params}. Right: 2D histograms of the ratio between estimated uncertainty and the deviation, in $\log_{10}$ scale. The parameter ranges on the vertical axes are the same as in the left panel, and the data only include the ``successful" population.}
        \label{fig: Internal Paramaters}
    \end{figure*} 
    
    \PIPS is shipped with optimised internal variables (for detail, see, Sec.~\ref{sec: algorithm}). These variables --- which we detail below --- are $N_\text{terms}$ $N_\text{peak}$, $R_\text{peak}$, and $N_0$, and each can be set when calling the fully automatic period-detection function \texttt{get\_period} in \PIPS. The effect each of these internal variables has on the performance of \PIPS is tested similarly to Sec.~\ref{sec:test_obs_params}, and we choose the default values we employ in \PIPS based on our observation.
    Using the method described in Sec.~\ref{sec: data construction}, data are generated from 20 different initial stars to have values of $N=35$, DBO = 7, DWR = 8, SNR = 3, the effects of which are discussed in Sec.~\ref{sec:test_obs_params}.
    
        \paragraph*{$N_{\rm terms}$} --- 
        $N_{\rm terms}$ determines the number of terms in the Fourier series \PIPS uses to approximate a star's light curve. In addition to affecting relative deviation and success rate, $N_{\rm terms}$ determines the lower limit on the number of observations needed for \PIPS to be able to run. When $N_{\rm terms} = 1$, \PIPS is nearly identical to LS. Based on the results shown in Fig.~\ref{fig: Internal Paramaters}, the ideal value of $N_{\rm terms}$ is either 4 or 5, with 5 providing a slightly larger success rate and 4 providing a slightly smaller deviation size\rev{, for the characteristics of data we fed to PIPS in this test.}
            
        \paragraph*{$R_\text{peak}$} ---
        $R_\text{peak}$ determines the number of samples taken when \PIPS closely investigates a peak. The only drawback to increasing this value is an increase in the computation time. The default value for $R_\text{peak}$ is 500, but if computation time is a concern, it can be lowered to 250 with minimal quality degradation.
            
        \paragraph*{$N_\text{peak}$} --- 
        $N_\text{peak}$ determines the number of peaks to closely investigate after \PIPS creates an initial periodogram. As with $R_\text{peak}$, the only drawback to increasing this value is more computation time. Increasing this value slightly improves the success rate, but has no effect on relative deviation.
            
        \paragraph*{$N_0$} ---
        \rev{$N_0$ refers to the number of points sampled in the periodogram for each peak.} Increasing $N_0$ slightly increases the success rate and decreases the relative deviation. The only drawback to increasing this value is more computation time.

\subsection{Accuracy of Reported Uncertainties}
    \PIPS is capable of reporting uncertainties that closely match the actual deviation from the true value of the period. When compared, the reported uncertainty and deviation from the true period are within one order of magnitude of each other 92\% $\pm$ 1\% of the time, as seen in Fig.~\ref{fig: All Paramaters}. This does not appear to hold true when $N < 25$, or when SNR $< 1.5$, but is otherwise true regardless of parameters.
    
\subsection{Performance by Star Subtypes}
    Of the RR Lyrae star subtypes tested, \PIPS performs better on RRc stars than on RRab, as the light curves of RRc stars are closer to sinusoids. Owing to the multimodal nature of RRd stars, data could not be generated in the way described above, so they were not used for validation. Cepheid variable stars in only the fundamental mode can be studied as well --- although without any adjustments, we see an increased deviation size. Results for each star type can be found in Fig.~\ref{fig: PIPS_Ceph_RRa_RRc} for reference.
    
\section{Additional Modules for Application}
    \label{sec:additional-modules}
    Beyond the period detection, we provide a wide variety of modules to enable a smooth workflow in the scientific analysis of periodically varying light curves. In the current version (\texttt{PIPS-0.3.0}), we have focused on pulsating variable stars. Below is a list of implemented modules.

    \subsection{Classification}
    \PIPS provides a set of tools related to the type-classifier for pulsating variable stars. This is the first step in our plan to implement a wide variety of classifiers with different methods, and we select a simple machine-learning algorithm and its pretrained classifier for the current version. We used OGLE \citep[][]{udalski2008ogle_iii,udalski2015_ogleiv,Soszynski_2009_OGLE_LMC,Soszynski2011_OGLE_BLG} and {\it Gaia} \citep[][]{gaia2016_main,gaia2018dr2,Eyer2017_GaiaDR1_var,GaiaDR2_variability} for training and validation, achieving 99\% training accuracy and 94\% testing accuracy. 

    The classifier included in \PIPS is based on the random forest (RF) classification algorithm \citep{Ho1995_RF}, which itself relies on the decision tree (D-Tree) algorithm. D-Trees and RFs have been used for numerous classification tasks on various datasets with considerable success \citep{Hinners2018_MLexoplanet,Ball2010_MLreview}. 
    The D-Tree algorithm works by hierarchically splitting the data, based on the given features, into subsets for each class such that each subset is maximally mutually dissimilar, while achieving most similarity within each group. This criterion of similarity is quantified by metrics such as the Gini index and entropy, which are the cost functions used to optimise the classifier \citep{derosa2016_gini}.
    The RF algorithm is an improvement upon the D-Tree classifier that randomly resamples the training data and creates multiple decision trees for these samples in a process known as bootstrapping. Each D-Tree uses a different, randomly chosen hierarchy of feature nodes (i.e., order of splitting) and makes its own decision on the classification. In the end, all the trees vote on the outcome, and the class with a plurality of votes is assigned to the data being classified \citep{Sarica2017_MLpubmed}.

    We use the RF classifier implemented in the \texttt{scikit-learn} Python package \citep[][]{scikit-learn}. 121,907 stars in total (10,281 classical Cepheids, 1856 Type~II Cepheids, 88,411 RRab stars, and 21,359 RRc stars) are extracted from the OGLE variable-star database \citep{udalski2015_ogleiv,udalski2008ogle_iii} to \rev{train} the algorithm. 900 stars from each class are then chosen as our training data, and the rest are earmarked as testing data to determine the accuracy of the model on the preclassified data. This is done to ensure that no bias is introduced owing to any mismatch in the sizes of the training data of the different classes.
    The OGLE database contains data on the stars' periods and the Fourier parameters. The model is trained on these data using the period and the $r_{21}$, $r_{31}$, $\phi_{21}$, and $\phi_{31}$ parameters. A 100-tree RF is used with a depth of 10. The model returns 99\% training accuracy. Upon deploying this model on the held-out testing data, the model returns a 98\%  accuracy score.

    To further test the validity of this model, we use the Gaia variable-star database \citep{gaia2016_main,gaia2018dr2,GaiaDR2_variability} as an additional testing set. Since these data were obtained from a completely different, independently compiled source, a good test score on them would make a strong case that the model is not biased toward the OGLE data sample used to train it. After removing objects that have missing information, we obtain 271 classical Cepheids, 43 Type~II Cepheids, 910 RRab stars, and 445 RRc stars in our {\it Gaia} testing set. Upon running the algorithm on these data, we obtain a 94\% testing accuracy, suggesting that the model is not biased toward OGLE data and is indeed applicable across datasets.
    
    \revv{Since the light-curve shape and amplitude vary between different passbands, a more accurate classification may be achieved by modeling such variations and including the passband information in the training. We plan to include this update in our next work that utilises multiband observation of RR Lyraes (Jennings et al., in prep).}

    \subsection{Phase-Offset Analysis}
    We implement a supplementary module to \texttt{PIPS} to calculate phase offsets (\textbf{O}bserved $-$ \textbf{C}alculated; $O-C$) over years of observations. Following \cite{Jurcsik2001_OC_OmCent,Jurcsik2012_OC_M3} and \cite{Szeidl2011_OC_M5}, we define the $O-C$ value to be
    \begin{equation}
        \label{eq:OC}
        (O-C)_i = \mu_i\ ,
    \end{equation}
    where $\mu_i$ is optimised so that
    \begin{equation}
        \label{eq:OC_mu}
        \mu_i = \text{argmin} \sum_j \frac{[\mathcal{F}(x_{ij} - \mu_i; P_{\textrm{mean}}, \theta_{\textrm{template}}) + b_i - y_{ij}]}{\sigma_{y,ij}}\ .
    \end{equation}
    Note that the dataset-unique offset $b_i$ is added to relax the possible differences in the calibration of telescopes. The template light-curve parameters are defined as the best-fit light curve at the reference period:
    \begin{multline}
        \label{eq:template}
        \theta_\text{template} = \\
        \text{argmin}\sum_j \frac{[\mathcal{F}(x_{\textrm{template},j};P_{\textrm{template}}, \theta_{\textrm{template}}) - y_{\textrm{template},j}]^2}{\sigma^2_{y, \textrm{template}, j}}\ .
    \end{multline}
    Here, $\mathcal{F}$ is the Fourier series and $\theta$ are Fourier parameters for the template light curve, $j$ denotes each exposure data point, and $i$ denotes the set of data for a historical dataset (see Jennings et al. 2021, in prep.).

    \subsection{Temperature Estimation}
    \begin{figure}
        \centering
        \includegraphics[width=\linewidth]{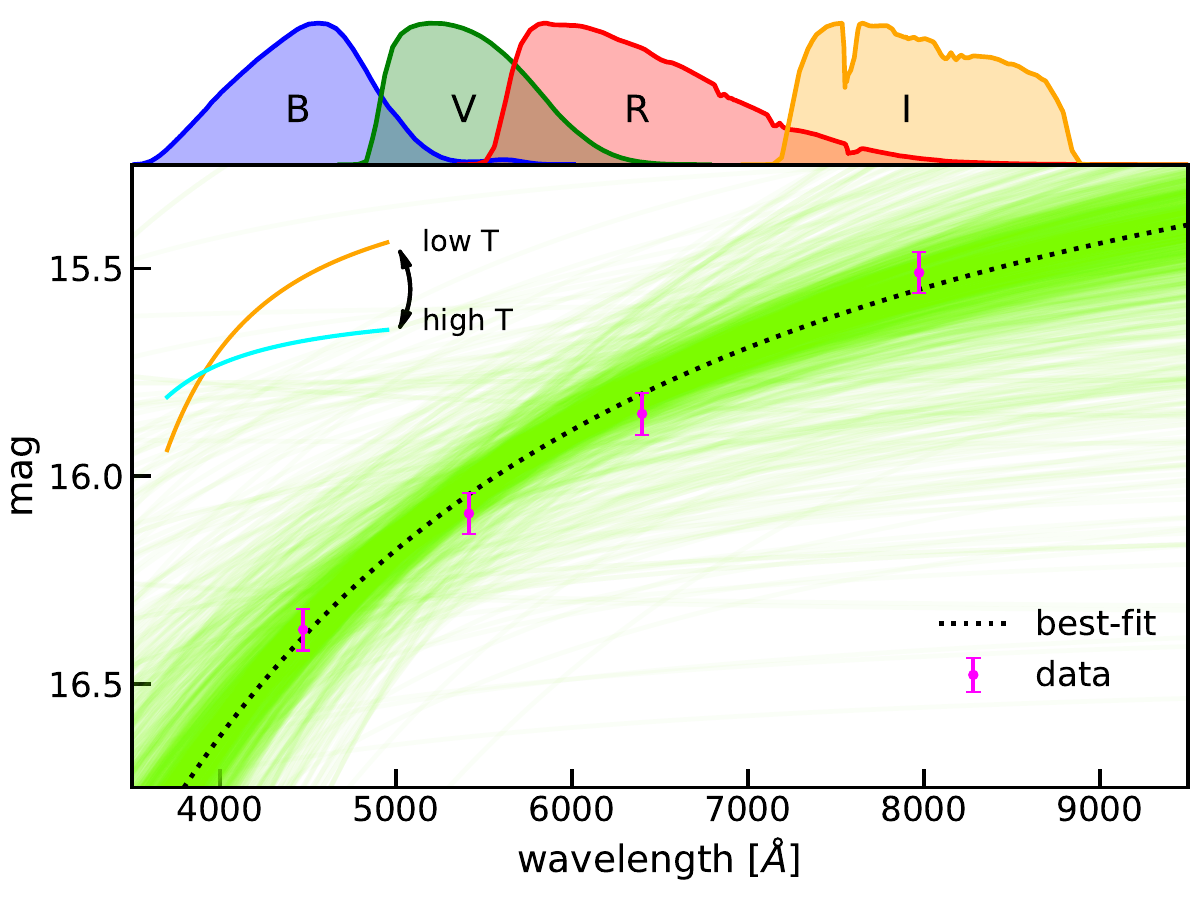}
        \caption{A visualisation of the temperature-estimation method in \PIPS. The Vega-calibrated flux (light-green curves; dotted black for the most probable case) is being fitted to the observed multiband photometric data (pink points) using the response function specific to the telescope (coloured curves at the top). The posterior analysis yields the fitted parameters of $T=6078\pm279$\,K and $b=14.2\pm0.2$\,mag. The intrinsic scatter $\sigma_\text{int}$ is consistent with zero, with the 68$^{\rm th}$ percentile at 0.13\,mag.}
        \label{fig:Temp_MCMC}
    \end{figure}

    When studying pulsating variable stars, the temperature is often regarded as a model-dependent value to be derived from other observables (see Sec.~\ref{sec: stellar properties}). However, statistical methods are capable of inferring the temperature from multiband photometry at a given snapshot. This adds another highly beneficial quantity to the observed values, possibly enabling more precise model constructions. In \PIPS, we provide a module that performs temperature-dependent spectral energy density (SED) optimisation for any given multiband photometry to create a time series of temperature values. 

    This analysis is performed under the assumption that the stellar flux can be approximated as blackbody radiation. In wavelength space, Planck's law,
    \begin{equation}
        \label{eq:BlackBody}
        F(\lambda,T) = \frac{2hc^2}{\lambda^5}\frac{1}{\exp\left(\frac{hc}{\lambda k_B T}\right)-1},
    \end{equation}
    gives us the estimated flux for a given temperature. For Vega-calibrated objects, the magnitude in any given band with the response function $\phi_\text{band}(\lambda)$ then becomes
    \begin{equation}
        \label{eq:Vega_MagSim}
        m_\text{sim}(T,b,\text{band}) = -2.5 \log_{10}\left(
        \frac{
        \int \phi_\text{band}(\lambda)F(\lambda,T)\mathrm d\lambda}
        {
        \int \phi_\text{band}(\lambda)F(\lambda,T_\text{Vega})\mathrm d\lambda}\right) + b\ .
    \end{equation}
    Here, the temperature for Vega \citep{Kinman2002_VegaTemp} is a fixed value, and the object's temperature $T$ and the constant offset\footnote{This constant offset $b$ absorbs the distance modulus and thus does not require an absolute magnitude in the fitting process.} vary.
    Using this $m_\text{sim}$ as simulated magnitudes, \PIPS performs MCMC to find the best-fit temperature, according to the maximum-likelihood function
    \begin{multline}
        \label{eq:Temp_MaxLikelihood}
        p(T,b,\sigma_\text{int}\mid\text{band},m,\sigma_m)\propto\\
        p(T,b,\sigma_\text{int})\ p(m\mid T,b,\text{band},\sigma_m,\sigma_\text{int})\ .
    \end{multline}
    In the equation above, $m$ is the observed apparent magnitude of the object in a given band, $\sigma_m$ is the uncertainty of the magnitude, and $\sigma_\text{int}$ is the intrinsic scatter for the magnitudes across all given bands. 
    We choose wide, uniform priors for the fitted parameters. For instance, the following are used for RRab stars at $\sim 16$\,mag:
    \begin{eqnarray}\nonumber
    T &\sim& \text{Uniform}(2000,10000)\\\nonumber
    b &\sim& \text{Uniform}(0,30)\\\nonumber
    \sigma_\text{int} &\sim& \text{HalfNormal}(1)\ .
    \end{eqnarray}
    A visualised example\footnote{For this example, we use four-band ($B$, $V$, $R$, and $I$) photometry data for RR Lyrae stars in globular cluster M3 (Jennings et al. 2022, in prep.), since the M15 data of \cite{H&M2020} are limited to two bands ($B$ and $V$). We utilised the Katzman Automatic Imaging Telescope (KAIT) at Lick Observatory \citep{Filippenko2001_KAIT} for the M3 observations, and the response functions are obtained from \cite{Stahl_2019}.} of this process is shown in Fig.~\ref{fig:Temp_MCMC}. 

    \subsection{Stellar Properties}
    \label{sec: stellar properties}
    \rev{ 
    For many objects (including RR Lyrae stars), the light-curve parameters recovered from the period-determination process can be combined with theoretical models to extract stellar properties such as mass, luminosity, colour, and metallicity \citep{kovacs2001empirical}. \PIPS implements a stellar parameter module that is integrated with its standard period-determination capabilities. As with period determination, this module is designed to accept a user-specified model as a flexible workflow, enabled by a robust and automatic error-propagation chain.}
    
    \rev{
    While the choice of model is intended to be open to the user, we have also implemented several stellar-parameter models as described by \citet{kovacs1994triple}, \cite{jurcsik1998fundamental}, and \citet{Cacciari_2005}. A point of concern that we intend to address with \PIPS is the lack of clear uncertainty treatment in any of these previous works. Perhaps because of this, we find that our results suggest that the current models may be somewhat inaccurate and require refinement. For further discussion, see Jennings et al. (2022, in prep.).}

\section{Discussion}
\label{sec:discussion}
    \subsection{False-Alarm Probability}
        In \PIPS, we do not implement  method to estimate the false-alarm probability (FAP). In the literature, the FAP has been considered to be an effective method to distinguish between true signal and false positives \citep{VanderPlas_2018}. 
        It is known, however, that the analytic form of the FAP is only available for the classical LS periodogram (i.e., a Fourier model with a single term); models other than that require bootstrapping to compute the FAP. Since we provide a more statistically robust interpretation of the periodogram (likelihood periodogram) and the statistical significance can be directly derived from it (Sec.~\ref{sec:likelihood}), we consider our solution to be more fundamental.

    \subsection{Planned Development}
        \paragraph*{Visualisation} --- Currently, our team is beginning development of a visualisation module that will allow users to explore the functionality of \PIPS. These visualisation tools will serve as a more intuitive way for users to utilise and interact with \PIPS with a minimal amount of extra code. Furthermore, we hope it will aid users in understanding how \PIPS works, enabling them to determine if the code is working properly and confirming that the outputs make sense. Such visualisation will also aid in user-input debugging, enabling users to understand why they might not be arriving at the results they are expecting. This module will initially make use of the \texttt{Ipython.widgets}\footnote{\url{https://ipywidgets.readthedocs.io/en/latest/}} library for Python, but it may eventually evolve into a custom GUI for \PIPS. Initial visualisation tools will allow users to explore the periodogram, light-curve folding, and $N_{\rm term}$ fitting process. Further implementations will potentially include a GUI allowing for the importing of data files and the setting of initial parameters via text box and buttons rather than lines of code.  
    
        \paragraph*{Exoplanets}--- A long-term goal for this codebase is for it to maintain a flexible collection of light-curve models, suitable for a variety of periodic fitting needs. One possible extension could be with respect to exoplanets --- e.g., by implementing the analytic light-curve models of \cite{mandel2002analytic} to take the place of our Fourier and Gaussian methods. 
        As shown in Section \ref{sec:model choice}, more appropriate models lead to better fits to a given light curve's period by a number of metrics. Therefore, creating an exoplanet-specific model may lead to further improvement over applying the standard \PIPS Fourier approach to exoplanet data. Rigorous testing against standards in the field such as box least squares \citep{kovacs2002box, kovacs2016bls}, transit least squares \citep{TLS_paper}, or wavelet-based methods \citep{regulo2007trufas} would be required to understand where in parameter space \PIPS could serve as an effective and useful tool. Comparative studies of transit-finding algorithms indicate that different approaches tend to complement one another \citep{moutou2005comparative}, so we expect even in the best-case scenario that \PIPS would serve as an addition to an existing set of tools available to exoplanet researchers.

    \subsection{Application to LSST}
        While the finalised plan for the Rubin Observatory Legacy Survey of Space and Time (LSST) is expected to be released later this year, we use information provided by \cite{Marshall_2017} about the baseline observing strategy (minion\_1016) to assess it as an application for \PIPS. LSST plans to take a large number of observations (199 in the $r$ band and 201 in the $i$ band) of every point in the southern sky over 10 years. Using the definitions from Sec.~\ref{sec:test_obs_params}, $N = 199$ and DBO = 18.35 for the $r$ band. The periodogram purity function, a measure of the phase completeness, varies considerably depending on location in the sky. DWR is related to phase completeness, indicating that DWR will be well suited to \PIPS in some regions, but not in others. For variable stars close enough to have high SNR and that occupy regions with good phase coverage, \PIPS can be used to detect the periods of stars from LSST data. Since the value of $N$ is so large for the $r$ and $i$ bands, it will likely be possible to divide observations into 4 or 5 subsets based on when they were taken for $N$ values of $\sim 50$ or $\sim 40$ (respectively), which may allow \PIPS to detect period changes over the 10\,yr of LSST observations.
        
\section{Conclusion}
\label{sec: long-term}
    \rev{We have presented our new approach, the Fourier-Likelihood (FL) periodogram, for fast and statistically robust period detection in astronomical time-series data. We have shown that, in many cases, the FL periodogram unifies the currently separate processes of period detection, uncertainty estimation, light-curve fitting, and statistical significance estimation, without the help of additional, computationally expensive calculations. In particular, the improved statistical significance estimation provides opportunities for a fast and efficient search for periodic signals in large datasets.}
    
    To enhance the advantage of the FL periodogram, we have developed our new \texttt{Python} package, \PIPS, which is designed to be an advanced platform for analysis tools of astronomical time-series data. The algorithm and mathematical/statistical models implemented in \PIPS were discussed, with primary focus on the period detection with FL periodogram. The main features include
    \begin{itemize}
        \item unlimited light-curve models, 
        \item automatic period detection,
        \item uncertainty estimation,
        \item \rev{various representations of periodogram (including FL)}, 
        \item statistical significance calculation, and
        \item automatic multiperiod detection.
    \end{itemize}
    The performance of \PIPS for period detection was tested with artificially generated data based on the true light curves, and we discussed the optimal range of data attributes and internal parameters that make the performance of \PIPS valid for scientific analysis.
    
    \PIPS also includes other modules, such as
    \begin{itemize}
        \item stellar parameter model helper,
        \item type classification,
        \item $O-C$ (phase offset) analysis tools, and
        \item multiband temperature estimation.
    \end{itemize}
    \rev{The detailed use of these modules and  scientific results for analysis of RR Lyrae stars will be presented in our forthcoming paper (Jennings et al. 2022, in prep.).}
    
    We also realise that the generalised light-curve fitter in \PIPS is possibly capable of complementing the existing methods for other types of variable objects, such as exoplanets. We plan to discuss this application in the near future (Savel et al. 2022, in prep.). 
    
    In the era of all-sky surveys and space telescopes, we are certain that with its advanced features and user-friendly interface, \PIPS will serve as a critical platform in the analyses of variable astronomical objects. We plan to continue implementing and updating a variety of tools to enhance the user experience and widen the range of roles that \PIPS can play. 

\section*{Acknowledgments}
    \rev{We are grateful to the anonymous reviewer for providing critical and constructive suggestions that led to a substantial improvement of this paper.}
    We thank Daniel Weisz, Joshua Bloom, and Kareem El-Badry for stimulating our interest in period analysis that initiated this project and providing feedback during the early development of \PIPS.
    Furthermore, we thank Keto Zhang for continuously providing his knowledge of statistics. 
    \revv{Y.S.M. thanks D'Arcy Kenworthy for his comments on statistical methods.}
    WeiKang Zheng is acknowledged for his support in scheduling and conducting the observations. 
    Finally, Y.S.M. thanks Alex Ho for providing great support when remote work was needed.
    
    This research made use of Astropy,\footnote{http://www.astropy.org} a community-developed core Python package for astronomy \citep{astropy:2013, astropy:2018}. 
    Also, it used data from the European Space Agency (ESA) mission {\it Gaia} (\url{https://www.cosmos.esa.int/gaia}), processed by the {\it Gaia} Data Processing and Analysis Consortium (DPAC, \url{https://www.cosmos.esa.int/web/gaia/dpac/consortium}). Funding for the DPAC has been provided by national institutions, in particular the institutions participating in the {\it Gaia} Multilateral Agreement.

    We acknowledge generous support from Marc J. Staley (whose fellowship partly funded B.E.S. as a graduate student whilst contributing to this work), the Christopher R. Redlich Fund, the U.C. Berkeley Miller Institute for Basic Research in Science (in which A.V.F. was a Miller Senior Fellow), and many individual donors to A.V.F.'s group. We also thank the U.C. Berkeley student observers involved with the research group: Samantha Stegman, Julia Hestenes, Keto Zhang, Teagan Chapman, Matthew Chu, Asia deGraw, Romain Hardy, Evelyn Lu, Emily Ma, Emma McGinness, Shaunak Modak, Derek Perera, Druv Punjabi, Jackson Sipple, Kevin Tang, Sergiy Vasylyev, Jeremy Wayland, and Abel Yagubyan. 
    
    KAIT and its ongoing operation were made possible by donations from Sun Microsystems, Inc., the Hewlett-Packard Company, AutoScope Corporation, Lick Observatory, the National Science Foundation, the University of California, the Sylvia \& Jim Katzman Foundation, and the TABASGO Foundation. Research at Lick Observatory is partially supported by a generous gift from Google. We appreciate the expert assistance of the staff at Lick Observatory.
    
\section*{Data Availability}
    \PIPS is available at the public software repository \texttt{PyPI} (\url{https://pypi.org/project/astroPIPS/}), \texttt{GitHub} (\url{https://github.com/SterlingYM/astroPIPS}), or by request to the authors. For detailed instructions and tutorials, visit the online documentation (\url{https://pips.readthedocs.io/en/latest/})
    The raw data used in our validation will be shared upon request to authors.
    The photometry of RR Lyrae stars used in Fig.~\ref{fig:periodogram_demo} is available at \url{https://github.com/SterlingYM/M15data2020}.




\bibliographystyle{mnras}
\bibliography{main.bib}
\vfill



\appendix
    \section{Basic Usage}
\label{sec:PIPS_usage}
In tandem with this paper, we release the stable version of \PIPS (\texttt{v0.3.0}), which contains all of the features and modules described here. Most of the period-analysis features are implemented in the \texttt{photdata} class, which is to be initialised with photometric time-series data. 
In \cite{H&M2020}, for instance, we initialised the \texttt{photdata} object for each RR~Lyrae star, using \texttt{time} (time in MJD), \texttt{mag} (observed apparent magnitudes), and \texttt{mag\_err} (uncertainty in each magnitude). The minimal working example of period detection with \PIPS in this case takes the following form:
\begin{lstlisting}[language=Python]
import PIPS
star = PIPS.photdata([time, mag, mag_err])
P, P_err, Z = star.get_period(return_Z=True)
\end{lstlisting}
The returned values \texttt{P}, \texttt{P\_err} , and \texttt{Z} are the most probable period, period uncertainty, and statistical significance ($Z$-score), respectively. The function \texttt{get\_period()} is a direct implementation of the likelihood-periodogram method described in Sec.~\ref{sec:likelihood}. When the statistical significance is not needed, an additional argument \texttt{repr\_mode='chi2'} reverts the method back to the $\chi^2$ periodogram-based method (Sec.~\ref{sec:chi-square-periodogram}), which is slightly faster.
A variety of settings can be applied as optional arguments. Available options include, but are not limited to, period search range, light-curve model (Sec.~\ref{sec:models}), number of terms ($K_\text{max}$ in Eq.~\ref{eq: Fourier fitting} and Eq.~\ref{eq:GMM}), model-specific parameters (e.g., $p$ in Eq.~\ref{eq:supergaussian}), and sampling parameters (e.g., $N_0$, $R_\text{peak}$ in Sec.~\ref{eq: periodogram}).
An example of code to specify these items is given below.
\begin{lstlisting}[language=Python]
P, P_err = star.get_period(
            repr_mode='chi2',
            p_min = 0.1,
            p_max = 1.0,
            method = 'custom',
            model = 'Gaussian',
            Nterms = 1,
            p = 5,
            N0 = 15,
            R_peak = 500,
           )
\end{lstlisting}
Here, \texttt{method='custom'} is passed to specify that a nonlinear model (currently, anything except Fourier model) has been selected.

Furthermore, it is made easy in \PIPS to visualise the periodogram. For instance, the line of code
\begin{lstlisting}[language=Python]
star.periodogram(model='Fourier',Nterms=5).plot()
\end{lstlisting}
automatically plots the periodogram, while simply calling the \texttt{periodogram()} function returns arrays of periods and periodogram powers. By default, periodogram is in $\chi^2$ interpretation for easier representation, and passing an argument \texttt{repr\_mode='likelihood'} forces the likelihood interpretation to be used.

Fore more detailed usage, including methods to implement custom light-curve models, visit the online documentation at  \url{https://pips.readthedocs.io/en/latest/}.
    \onecolumn 
    \section{Marginalisation of Bivariate Gaussian}
\label{sec:appendix_gaussian}
\rev{
Here we provide a more detailed derivation of Eq.~\ref{eq:bivariate_slice}. To begin, we define $\tilde\mu_2$, a function of $\theta_1$ along the best-fit line to the distribution
\begin{eqnarray}
    \tilde\mu_2 = \mu_2 + \rho\frac{\sigma_2}{\sigma_1}\left(\theta_1-\mu_1\right)\ ,
\end{eqnarray}
where $\mu$ and $\sigma$ are the mean and standard deviation values of parameters, respectively, and $\rho$ denotes the correlation between two parameters, $\theta_1$ and $\theta_2$. Then, 
\begin{eqnarray}
    \label{eq:bivariate_normal}\nonumber
    \mathcal{L}(\theta_2\given \theta_1, x,y,\sigma_y)
    &=& \frac{1}{2\pi\sigma_1\sigma_2\sqrt{1-\rho^2}}\exp\left(
        \frac{-1}{2(1-\rho^2)}\left[
            \left(\frac{\theta_1-\mu_1}{\sigma_1}\right)^2
            -2\rho\left(\frac{\theta_1-\mu_1}{\sigma_1}\right)\left(\frac{\theta_2-\mu_2}{\sigma_2}\right)
            +\left(\frac{\theta_2-\mu_2}{\sigma_2}\right)^2
        \right]\right)\\\nonumber
    &=& \frac{1}{2\pi\sigma_1\sigma_2\sqrt{1-\rho^2}}\exp\left(
        \frac{-1}{2(1-\rho^2)}\left[
            \left(\frac{\theta_1-\mu_1}{\sigma_1}\right)^2
            -2\left(\frac{\tilde\mu_2 - \mu_2}{\sigma_2}\right)\left(\frac{\theta_2-\mu_2}{\sigma_2}\right)
            +\left(\frac{\theta_2-\mu_2}{\sigma_2}\right)^2
        \right]\right)\\\nonumber
    &=& \frac{1}{2\pi\sigma_1\sigma_2\sqrt{1-\rho^2}}\exp\left(
        \frac{-1}{2(1-\rho^2)}\left[
            \left(\frac{\theta_1-\mu_1}{\sigma_1}\right)^2
            - \rho^2\left(\frac{\theta_1-\mu_1}{\sigma_1}\right)^2 
            + \left(\frac{\theta_2-\tilde\mu_2}{\sigma_2}\right)^2
        \right]\right)\\\nonumber
    &=& \frac{1}{2\pi\sigma_1\sigma_2\sqrt{1-\rho^2}}\exp\left[-
        \frac{(1-\rho^2)}{2(1-\rho^2)}\left(\frac{\theta_1-\mu_1}{\sigma_1}\right)^2\right]
        \exp\left[-\frac{\left(\theta_2-\tilde\mu_2\right)^2}{2\left(1-\rho^2\right)\sigma_2^2}\right]\\
    &\equiv& \alpha(\theta_1)\exp\left[-\frac{\left(\theta_2-\tilde\mu_2\right)^2}{2\beta^2}\right]\ ,
\end{eqnarray}
This is a slice of the multivariate Gaussian at fixed $\theta_1$ along the $\theta_2$ direction. 
The height of the Gaussian ($\alpha$) is a function of $\theta_1$, and the standard deviation $\beta$ is determined by the variance-covariance matrix only (i.e., independent of $\theta_1$). This suggests that 
\begin{eqnarray}
    \label{eq:bivariate_integral}\nonumber
    && \int\mathcal{L}(\theta_1,\theta_2\given x,y,\sigma_y)\ \mathrm d\theta_2 \\
    && \quad =
    \alpha(\theta_1) \int \exp\left(\frac{-\left[\theta_2-\mu_2(\theta_1)\right]^2}{2\beta^2}\right)\ \mathrm d\theta_2 \\
    && \quad \equiv
    \alpha(\theta_1) \cdot I(\theta_2-\mu_2)\ ,
\end{eqnarray}
and the integral term $I$ becomes independent of the exact value of $\theta_1$ and $\mu_1$.
}
\newpage

\section{Supplemental figures}
\begin{figure}
    \centering
    \includegraphics[width=\linewidth]{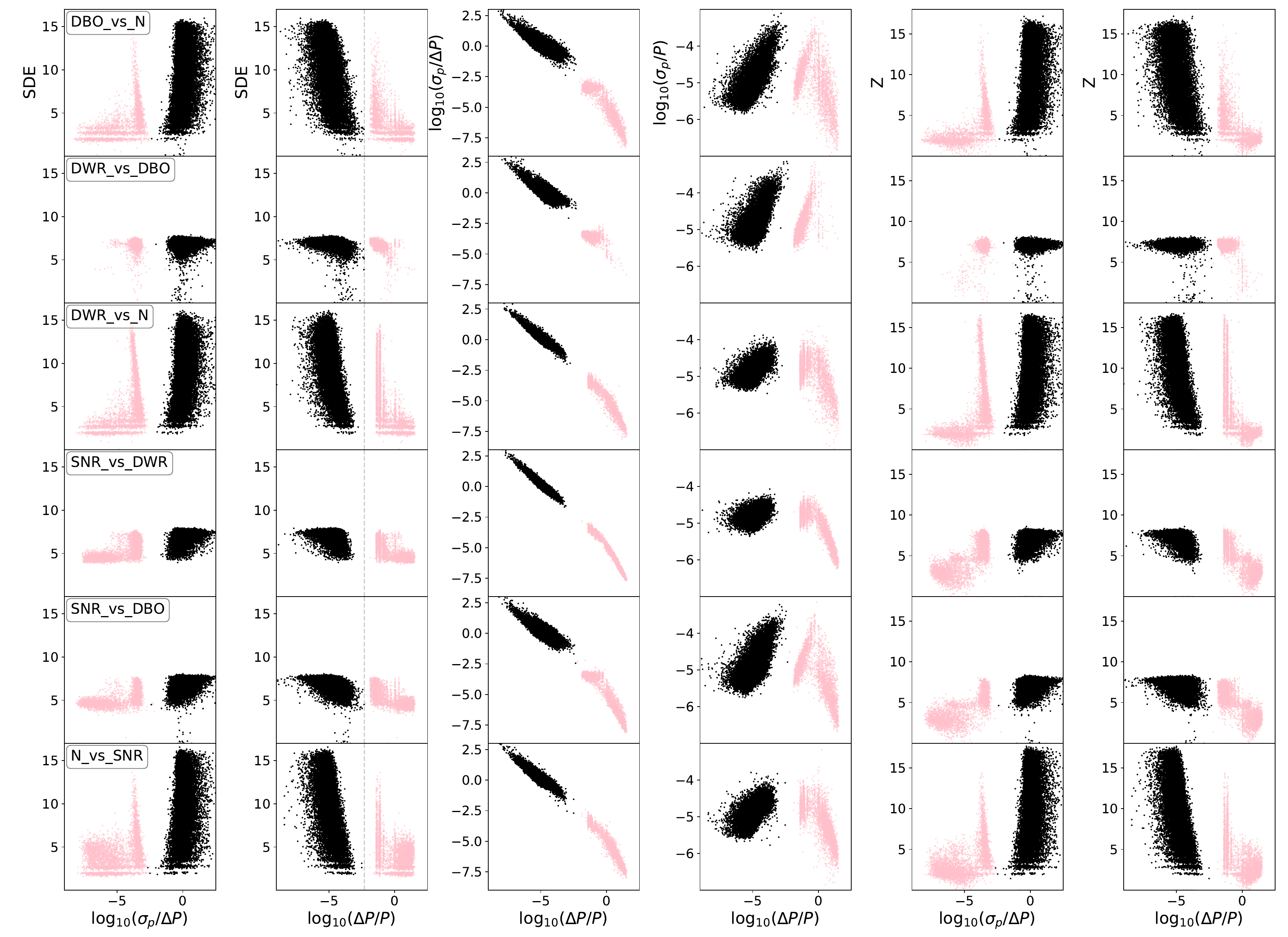}
    \caption{Distribution of the ``successful'' (black) and ``unsuccessful'' (light pink) populations in various parameters. Each row corresponds to the results in each panel in Fig.~\ref{fig: All Paramaters}. It is evident that there is a clear separation at $\log_{10}(\Delta P/P)\approx -2.3$ (grey dashed line), which corresponds to $0.5\%$ error. This results in clear separations in all other visualised parameters, such as $\log_{10}(\sigma_P/P)$. The statistical metrics, SDE and $Z$-value, show nearly identical distributions for the successful populations, while the typical $Z$-values for unsuccessful populations tend to be smaller than SDE. This indicates that the likelihood-based $Z$-value is a superior metric to evaluate the significance of the detected period.}
    \label{fig:separation_plot}
\end{figure}

\begin{figure}
    \centering
    \includegraphics[width=\linewidth]{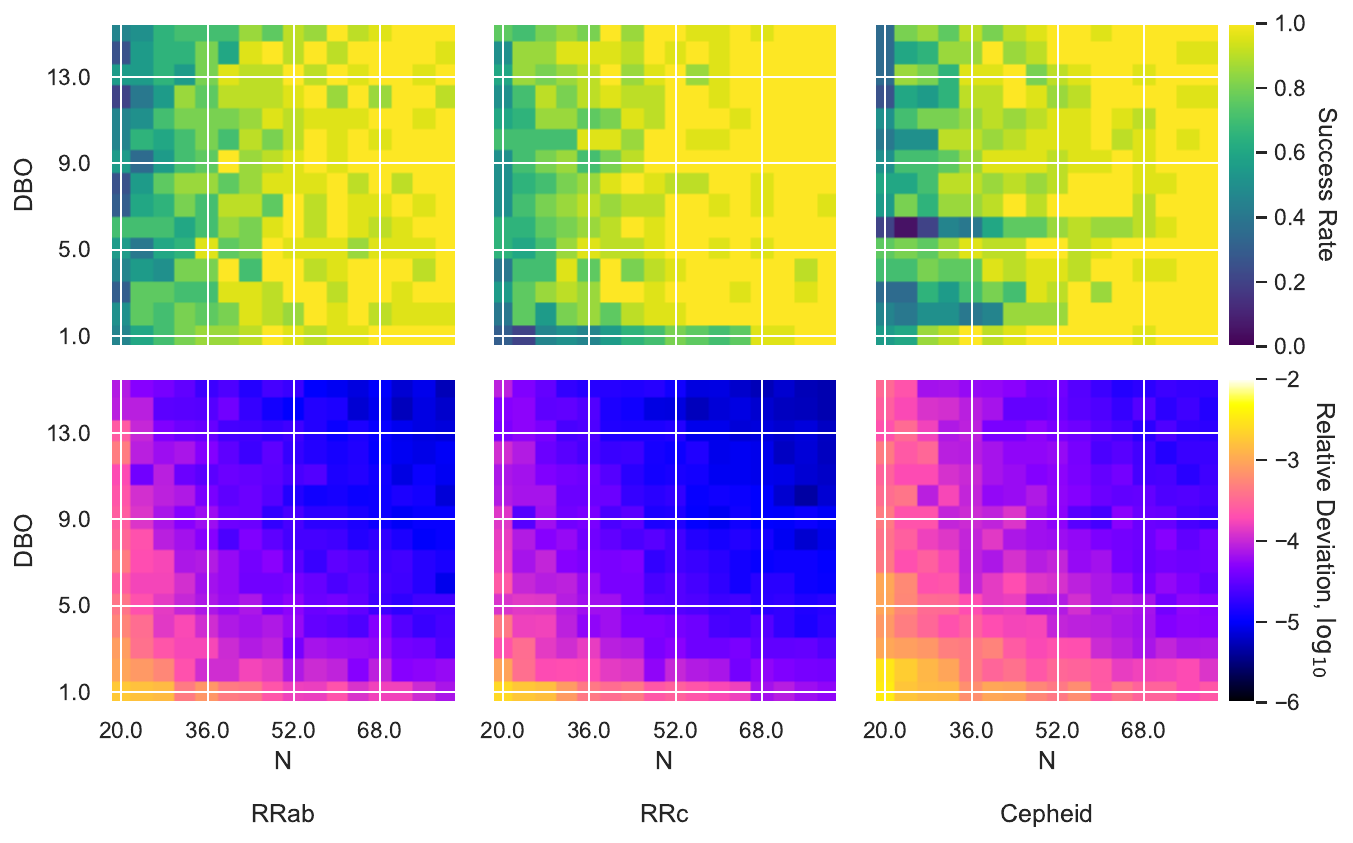}
    \caption{The success rate (top panels) and relative deviation from the known period (bottom panels) with \PIPS data generated using fundamental mode Cepheids at given number of observations ($N$) and DNO, in comparison to RRab and RRc stars. SNR = 5 and \winrat{} = 8 are kept constant. It is apparent that \PIPS yields the better results in both the success rate and the relative deviation for RRc stars. The relative deviation for Cepheids suggests that \PIPS requires larger $N$ and DBO to achieve results similar to those of RRab/c stars, in contrast.}
    \label{fig: PIPS_Ceph_RRa_RRc}
\end{figure}

\bsp	
\label{lastpage}
\end{document}